\documentclass[%
 reprint,
 amsmath,amssymb,
 $aps,
prb,
]{revtex4-1}

\usepackage{graphicx} % Include figure files
\usepackage{bm}
\usepackage{txfonts}
\usepackage{mathrsfs}
\usepackage{amsmath}
\usepackage{xcolor}

\usepackage{subcaption}

\def\b0{{\mathbf 0}}

\def\b0{{\mathbf 0}}

\def\beq{\begin{equation}}
\def\eeq{\end{equation}}
\def\beqa\begin{eqnarray}
\def\eeqa{\end{eqnarray}}

\begin{document}

%\title{Nature of the superfluid quantum phase transition in imbalanced Fermi mixtures} 
\title{Quantum Lifshitz points and fluctuation-induced first-order phase transitions in imbalanced Fermi mixtures}
\author{Piotr Zdybel}
\affiliation{Institute of Theoretical Physics, Faculty of Physics, University of Warsaw, Pasteura 5, 02-093 Warsaw, Poland}
%\affiliation{Max-Planck-Institute for Solid State Research, Heisenbergstr.\ 1, D-70569 Stuttgart, Germany}
%
\author{Pawel Jakubczyk }
\affiliation{Institute of Theoretical Physics, Faculty of Physics, University of Warsaw, Pasteura 5, 02-093 Warsaw, Poland} 
\date{\today}
\begin{abstract}
We perform a detailed analysis of the phase transition between the uniform superfluid and normal phases in spin- and mass-imbalanced Fermi mixtures. At mean-field level we demonstrate that at temperature $T\to 0$  the gradient term in the effective action can be tuned to zero for experimentally relevant sets of parameters, thus providing an avenue to realize a quantum Lifshitz point. We subsequently analyze  damping processes affecting the order-parameter field across the phase transition. 
%We resolve the structure of the effective action and analyze the excitation spectrum in the vicinity of the phase boundary.  
We show that, in the low energy limit, Landau damping occurs only in the symmetry-broken phase and affects exclusively the longitudinal component of the order-parameter field. It is however unavoidably present in the immediate vicinity of the phase transition at temperature $T=0$. We subsequently perform a renormalization-group analysis of the system  in a situation, where, at mean-field level, the quantum phase transition is second order (and not multicritical). %In absence of Landau damping, we recover the expected quantum-classical crossover and quantum-critical behavior with the dynamical exponent $z=1$. 
We find 
%however 
that, at $T$ sufficiently low, including the Landau damping term in a form derived from the microscopic action destabilizes the renormalization group flow towards the Wilson-Fisher fixed point. This signals a possible tendency to drive the transition weakly first-order by the coupling between the order-parameter fluctuations and fermionic excitations effectively captured by the Landau damping contribution to the order-parameter action.

\end{abstract}

\pacs{}

\maketitle

%%%%%%%%%%%%%%%%%%%%%%%%%%%%%%%%%%%%%%%%%%%%%%%%%%%%%%%%%%%%%%%%%%%%%%%%%%

%%%%%%%%%%%%%%%%%%%%%%%%%%%%%%%%%%%%%%%%%
\section{Introduction}
%%%%%%%%%%%%%%%%%%%%%%%%%%%%%%%%%%%%%%%%%

Ultracold Fermi gases have been a topic of growing interest over the last years.\cite{Radzihovsky_2007, Bloch_2008, Giorgini_2008,  Gubbels_2013, Gross_2017, Strinati_2018, Cooper_2019} This is related to the %experimentally accessible 
rich phenomenology involving the interplay of quantum statistics and interaction effects, including aspects hardly reachable or controllable in traditional electronic systems. A standard example is the BEC-BCS crossover,\cite{BCS_BEC_book, Randeria_2014} experimentally realized by resonantly controlled $s$-wave interaction strength between particles in different pseudospin states.
In Fermi mixtures, the possibility of tuning the  population\cite{Zwierlein_2006, Partridge_2006, Ketterle_2009} and mass imbalance\cite{Wille_2008, Tiecke_2010, Trenkwalder_2011, Jag_2014, Ravensbergen_2018, Ravensbergen_2020, Neri_2020, Hara_2011} allows for studying unconventional pairing phenomena leading to unusual superfluid phases  such as  the breached pair (Sarma-Liu-Wilczek) superfluids\cite{Sarma_1963, Liu_2003} or the Fulde-Ferrell-Larkin-Ovchinnikov (FFLO) phases.\cite{fulde_superconductivity_1964, larkin_nonuniform_1965}

%Moreover, by using resonantly controlled an $s$-wave interaction strength between particles in different pseudospin states, the BEC-BCS crossover is attainable. The accompanying rapid development of the experimental techniques and ability to engineer Fermi-Fermi mixtures of ultracold atoms, such as $^{161}$Dy$-^{40}$K or $^6$Li$-^{40}$K mixtures, makes this issue a graceful topic of research.

Of substantial interest are quantum phase transitions between the superfluid and the normal, Fermi-liquid-like phases in imbalanced systems.\cite{Iskin_2007, Parish_Nat_2007, Radzihovsky_2010, Klimin_2012, Gubbels_2013, Strack_2014, Roscher_2015, Wang_2017} Increasing the  composition- (and/or mass-) imbalance mismatches the Fermi surfaces and  suppresses  pairing. The population and/or mass imbalance  therefore play the role of realistically controllable parameters to tune the system across the phase transition  both at temperature $T=0$ and $T>0$. 

An interesting, and, thus far, not fully explored aspect of the phase diagram of Fermi mixtures concerns multicritical phenomena. It is well recognized that, in a typical situation, the phase transition between the uniform superfluid and normal phases is first-order for $T$ sufficiently low, and necessarily becomes continuous for $T$ higher, above a tricritical temperature $T_{tri}$.\cite{Parish_2007, Lamacraft_2008, Chevy_2010} It has, however, recently been shown\cite{Zdybel_2018} that (in dimensionality $d=3$) $T_{tri}$ can actually be suppressed to zero for sufficiently large mass imbalance. In this situation, the phase diagram exhibits a quantum critical point (QCP). As was indicated by mean-field studies, there is however another, quite distinct multicritical phenomenon present at $T>0$, related to the so-called Lifshitz point,\cite{Gubbels_2009, baarsma_population_2010} where two ordered phases (uniform and nonuniform superfluids) coexist with the normal phase. The universal critical singularities at the classical ($T>0$) Lifshitz point\cite{Diehl_2002, Diehl_2003, Zappala_2018} are completely different from those controlling the critical or tricritical point. Notably, while the upper critical dimension $d_u$ for the tricritical transition is 3, it is way higher (at least 9/2) for the Lifshitz transition,\cite{Diehl_2002, Essafi_2012, Zappala_2017} making the conventional approaches to critical phenomena in spatial dimensionality $d=3$ based on the $\epsilon$-expansion ($\epsilon=d_u-d$)  problematic.\cite{Essafi_2012} The presence of a Lifshitz point additionally leads to rich and interesting crossover phenomena.\cite{Diehl_2002}

 Phase transitions at $T>0$ and $T=0$ are driven by thermal and quantum fluctuations respectively. The natural question concerning the physics of imbalanced Fermi gases is whether the Lifshitz point may be suppressed to zero temperature  and arise as a result of purely quantum fluctuations. In this paper we give a positive answer to this question, deriving an analytical mean-field criterion for the occurrence of such a quantum Lifshitz point  and providing a proposal for its potential experimental realization.   

%Realization of a quantum Lifshitz point represents an interesting situation which, however, requires fine-tuning. 

The second problem addressed in the present paper concerns the nature of the quantum phase transition between the normal and uniform superfluid phases in a situation when the transition is continuous (and not multicritical) at mean-field level. The general approach to quantum criticality in Fermi systems, based on an effective bosonic mode is recognized as the Hertz-Millis theory.\cite{Hertz_1974, Millis_1993} Within this approach, the interaction between the collective order-parameter mode and soft fermionic excitations across the Fermi surface is captured by the so-called Landau-damping term, appearing in the inverse propagator of the order-parameter mode. The presence (or absence) of Landau damping, as well as its precise form has profound impact on the quantum critical behavior, in particular it determines the value of the dynamical critical exponent $z$, which in turn influences many thermodynamic and transport properties. As was shown in Ref.~\onlinecite{Zdybel_2019},  Landau damping is always present in the ordered phase in the vicinity of the imbalance-driven superfluid quantum phase transition. We here supplement and extend the analysis of Ref.~\onlinecite{Zdybel_2019} by further elucidating the nature of Landau damping in the superfluid phase. In particular, we demonstrate that it affects exclusively the longitudinal component of the order-parameter field. By considering the normal phase, we show in turn that Landau damping is absent therein. This constitutes a situation different as compared to that studied for magnetic quantum critical points and requires a renormalization-group (RG) treatment formulated in the ordered phase. We address this problem within the framework of non-perturbative RG. Our results, based on a very simple truncation of a functional RG flow equation, recover the expected behavior of the system in absence of Landau damping. %The physics involved may be understood as a crossover phenomenon governed by two RG fixed points (Gaussian and Wilson-Fisher in our case). 
However, for temperatures low enough, the presence of  Landau damping in the form derived from the microscopic model turns out to destabilize the expected flow towards the Wilson-Fisher fixed point, so that the asymptotic scaling regime is not reached. This behavior may indicate a fluctuation-driven first-order transition induced by Landau damping and would constitute a mechanism for fluctuation driven first order phase transitions entirely different as compared to the one well-studied e.g. for the ferromagnetic quantum phase transition.\cite{Belitz_2005} 

The outline of the present paper is as follows: In Sec.~II, focusing on the symmetry-broken phase,  we summarize the structure of the effective order-parameter action employed in our study of the superfluid phase transition. Sec.~III contains an analysis of the pair fluctuation propagator in the normal phase, in particular the properties of the gradient expansion (including both analytical and non-analytical contributions). This leads to insights concerning order-parameter excitations as well as the dominant instabilities towards the uniform or non-uniform ordered phases. We derive a criterion for the occurrence of a quantum Lifshitz point and discuss its phenomenological implications. We subsequently discuss the nature of the (nonlocal) terms related to Landau damping and demonstrate their absence in the normal phase in the low-energy limit. 
In Sec.~IV we present a study of order parameter fluctuation effects focusing on a situation, where the quantum phase transition is second-order (and not multicritical) at mean-field level. We emphasize the necessity to formulate the RG in the symmetry-broken phase.\cite{Jakubczyk_2008} We discuss the effect of destabilizing the RG flow towards the Wilson-Fisher fixed point at temperatures ($T>0$) sufficiently low caused by the Landau damping. We summarize the paper in Sec.~V.

%%%%%%%%%%%%%%%%%%%%%%%%%%%%%%%%%%%%%%%%%
\section{Effective action}
%%%%%%%%%%%%%%%%%%%%%%%%%%%%%%%%%%%%%%%%%

We consider a two-component fermionic mixture of particles of masses $m^\sigma$ and chemical potentials $\mu^\sigma$ [$\sigma\in\{+,-\}$], where an attractive interspecies contact interaction $g<0$ triggers $s$-wave pairing.  The system is described by the microscopic action
\beq
S_\psi=\int_k \sum_{\sigma} \bar{\psi}^\sigma_k\left[-G_{0,\sigma}^{-1}(k)\right]\psi^\sigma_k +g\int_{k,k',q}\bar{\psi}^+_{k+q}\bar{\psi}^-_{-k}\psi^-_{k'+q}\psi^+_{-k'}
\label{Eq1}
\eeq
and the grand-canonical partition function reads
\beq
Z=\int \mathcal{D}[\bar{\psi}_k^\sigma, \psi_k^\sigma]\,\mathrm{e}^{-S_\psi}\;.
\label{Eq2}
\eeq
Here $\{\bar{\psi}_k^\sigma, \psi_k^\sigma\}$ are Grassmann fields, $G_{0,\sigma}(k)=(ik_0-\xi_k^\sigma)^{-1}$, $\xi_k^\sigma=\vec{k}^2/(2m^\sigma) -\mu^\sigma$ denotes the~dispersion relation  and $k=(k_0, \vec{k})$ collects the (fermionic) Matsubara frequency  $k_0=2\pi T(n+1/2)~[n\in\mathbb{Z}]$ and momentum $\vec{k}$. Throughout the paper we put $k_B=\hbar=1$ and use the shorthand notation $\int_k (\cdot)=T\sum_{k_0}\int_{\vec{k}}\,(\cdot)=T\sum_{k_0}\int \frac{\mathrm{d}^d \vec{k}}{(2\pi)^d} (\cdot)$.

The effective action $\mathcal{S}_0$ for the order-parameter field can be derived using a standard procedure described in detail in Ref.~\onlinecite{Iskin_2007}. We decouple the interaction term in Eq.~(\ref{Eq1}) into the Cooper channel via the Hubbard-Stratonovich transformation.\cite{Altland_book} This introduces the $s$-wave pairing field $\phi_q$ conjugate to the fermionic bilinear  $\bar{B}_q=\int_k \bar{\psi}^+_{k+q}\bar{\psi}^-_{-k}$. Subsequently  we integrate out the Grassmann fields $\{\bar{\psi}_k^\sigma, \psi_k^\sigma\}$, which leads to a purely bosonic description of the system. We obtain the following expression 
\beq
\mathcal{S}_0[\phi]=\frac{1}{2}\int_{q}\Phi^*_q\;\mathbb{F}^{-1}(q)\,\Phi_q +\mathcal{U}[\phi]\;,
\label{Eq3}
\eeq
where $\Phi_q^*=[\tilde{\phi}_q^*, \tilde{\phi}_{-q}]$,  $\Phi_q=[\tilde{\phi}_q , \tilde{\phi}^*_{-q}]^T$ and $\phi_q=\phi_0\delta_{q,0}+\tilde{\phi}_q$ is the bosonic pairing field split into the uniform contribution $\phi_0$ and fluctuation $\tilde{\phi}_q$  (notice that $\tilde{\phi}_{q=0}=0)$. The quantity $q$ encompasses the (bosonic) Matsubara frequency $q_0$ and momentum $\vec{q}$. All the momentum expansions will now be performed around $\vec{q}=0$ which restricts the results below to uniform phases. Nonuniform ordering tendencies are discussed in Sec.~III. 

The term $\mathcal{U}[\phi]$ may be written as 
\beq
\mathcal{U}[\phi]=\int_x U(\phi)\;,
\eeq
where $x=(\tau, \vec{x})$ collects the imaginary time and position vector, while $\int_x(\cdot)=\int_0^{1/T}\mathrm{d}\tau\int\mathrm{d}^d x\,(\cdot)$.
 The effective potential $U(\phi)$ takes the form\cite{Zdybel_2018}  
%corresponds to $\tilde{\Gam}[\phi]$ evaluated at a uniform configuration of the pairing field $\phi_0$, i.e. $\mathcal{U}[\phi_0]=\Gam[\phi_0]$. In our case, it has the following form\cite{Zdybel_2018} 
\beq
U(\phi)=-\frac{|\phi|^2}{g}+T\int_{\vec{k}}\sum_\sigma \ln f\left(-E_k^\sigma\right),
\label{Eq4}
\eeq
where $f(x)=[\exp(x/T)+1]^{-1}$ and 
\beq
E_k^\sigma=\frac{\xi_k^+-\xi_k^-}{2}+\sigma\sqrt{\left(\frac{\xi_k^++\xi_k^-}{2}\right)^2+|\phi|^2}
\label{Eq5}
\eeq
is the excitation energy of fermionic quasiparticles. The global minimum of $U(\phi)$  determines the mean-field grand-canonical potential per unit volume.

Of our major present interest is the inverse Gaussian pair fluctuation propagator matrix $\mathbb{F}^{-1}(q)$ in Eq.~(\ref{Eq3}), encoding  the dynamical characteristics of the collective excitations in the system.   Its properties in the superfluid phase were recently addressed in Ref.~\onlinecite{Zdybel_2019}, leading \emph{inter alia} to the conclusion that Landau damping of the Anderson-Bogolyubov mode is unavoidable close to the quantum phase transition approaching  from the superfluid phase. Damping brings about modification of the dynamics of the system, which is manifested by terms $\sim\frac{|q_0|}{|\vec{q}|}$ in a~gradient expansion of matrix elements of the inverse propagator $\mathbb{F}^{-1}(q)$. Denoting $M_{ij}(q):=[{F}^{-1}(q)]_{ij}$, we obtain\cite{Zdybel_2019}
\begin{align}
M_{11}(q)=M_{22}(-q)=\frac{1}{2}\left[Z\vec{q}^2-iWq_0+Z_0q_0^2+\frac{L\rho_0}{2}\frac{|q_0|}{|\vec{q}|}\right],
\label{Eq7}
\end{align}
\beq
M_{12}(q)=M_{21}(q)=\frac{\rho_0}{2}\left[Y\vec{q}^2+Y_0q_0^2+\frac{L}{2}\frac{|q_0|}{|\vec{q}|}\right],
\label{Eq8}
\eeq
where we parametrize the action in such a way that terms $M_{ij}(0)$ are contained in the local potential term $\mathcal{U}[\phi]$. The quantities $\{Z,Z_0,Y,Y_0,W,L\}$ may be related to the parameters of the microscopic fermionic action $S_\psi$, while $\rho_0:=|\phi_0|^2$. 
 It is noteworthy that contributions related to the damping of collective modes are proportional to $\rho_0\frac{|q_0|}{|\vec{q}|}$. In the superfluid phase, the quasiparticles are a superposition of particle-like and hole-like excitations.\cite{Pieri_2004, Strinati_2018} As a consequence, in the evaluation of the matrix elements of $\mathbb{F}^{-1}(q)$, both particle-particle and particle-hole bubble diagrams may give a contribution.\cite{Pieri_2004} Pair fluctuations $\tilde{\phi}_q$ couple with fermionic particle-hole excitations across the Fermi level, which gives rise to the Landau damping.\cite{Kurkjian_2017, Klimin_2019} However, once we move to the normal phase ($\rho_0=0$), the particle-hole continuum vanishes, and damping terms disappear.\cite{Zdybel_2019} This manifests itself by the proportionality of the damping terms to $\rho_0$. This feature makes the nature of the phase transition in  imbalanced Fermi mixtures distinct from those studied in the context of magnetic phase transitions (see e.g. Refs.~\onlinecite{Nagaosa_book, Continentino_book, Belitz_2005, Lohneysen_2007}).

We now decompose $\tilde{\phi}_q$  using the cartesian representation %\cite{Strack_2008, Obert_2013}
\beq
\tilde{\phi}_q=\sigma_q+i\pi_q,~~~~~\tilde{\phi}^*_q=\sigma_{-q}-i\pi_{-q},
\label{Eq8}
\eeq
where $\sigma_q$ and $\pi_q$ are bosonic fields describing longitudinal and transverse fluctuations of the pairing field, respectively. The uniform contribution $\phi_0$ is chosen real. We have $\sigma^*_{q}=\sigma_{-q}$ and  $\pi^*_{q}=\pi_{-q}$ since $\sigma(x)$ and $\pi(x)$ are real fields. Inserting the decomposition of Eq.~(\ref{Eq8})  into Eq.~(\ref{Eq3})  and parametrizing the effective potential via a quartic form 
\beq
U(\phi)= \frac{\lambda}{2}\left(\rho-\rho_0\right)^2\;,\;\;\;\;\; \rho:=|\phi|^2
\label{Eq6}
\eeq
leads to
\beq
\mathcal{S}_0=\mathcal{S}_{\sigma^2}+\mathcal{S}_{\pi^2}+\mathcal{S}_{\sigma\pi}+
\mathcal{S}_{\sigma^3}+\mathcal{S}_{\sigma\pi^2}+\mathcal{S}_{\sigma^4}+\mathcal{S}_{\pi^4}+\mathcal{S}_{\sigma^2\pi^2},
\label{Eq10}
\eeq
where the quadratic terms are obtained as 
\beq
\mathcal{S}_{\sigma^2}=\frac{1}{2}\int_q\left[m_\sigma^2+Z_\sigma^0 q_0^2+Z_\sigma\vec{q}^2+L\rho_0\frac{|q_0|}{|\vec{q}|}\right]\sigma_q\sigma_{-q},
\label{Eq11}
\eeq
\beq
\mathcal{S}_{\pi^2}=\frac{1}{2}\int_q\left[Z_\pi^0 q_0^2+Z_\pi\vec{q}^2\right]\pi_q\pi_{-q},
\label{Eq12}
\eeq
\beq
\mathcal{S}_{\sigma\pi}=-\frac{1}{2}\int_qWq_0\left[\sigma_q\pi_{-q}-\pi_q\sigma_{-q}\right]\;.
\label{Eq13}
\eeq
Here $m_\sigma^2=2\lambda\rho_0$ is a mass of the $\sigma$-field, $Z_\sigma=Z+\rho_0 Y$, $Z^0_\sigma=Z_0+\rho_0 Y_0$, $Z_\pi=Z-\rho_0 Y$ and $Z^0_\pi=Z_0-\rho_0 Y_0$. 

 For $L=0$ the obtained action $\mathcal{S}_0$ is reminiscent of the one studied for interacting bosons (see e.g. \onlinecite{Pistolesi_2004, Obert_2013}). As expected, the transverse $\pi$-mode is massless. We observe  that the Landau damping term occurs only in the longitudinal mode. We also note that its presence was not taken into account in previous studies of fluctuation effects in imbalanced Fermi systems (e.g. \onlinecite{Strack_2014, Boettcher_2015, Zdybel_2018}).
%%%%%%%%%%%%%%%%%%%%%%%%%%%%%%%%%%%%%%%%%
\section{Pair fluctuation propagator in the normal phase}
\begin{figure}[ht] 
\begin{center}
\includegraphics[width=7cm]{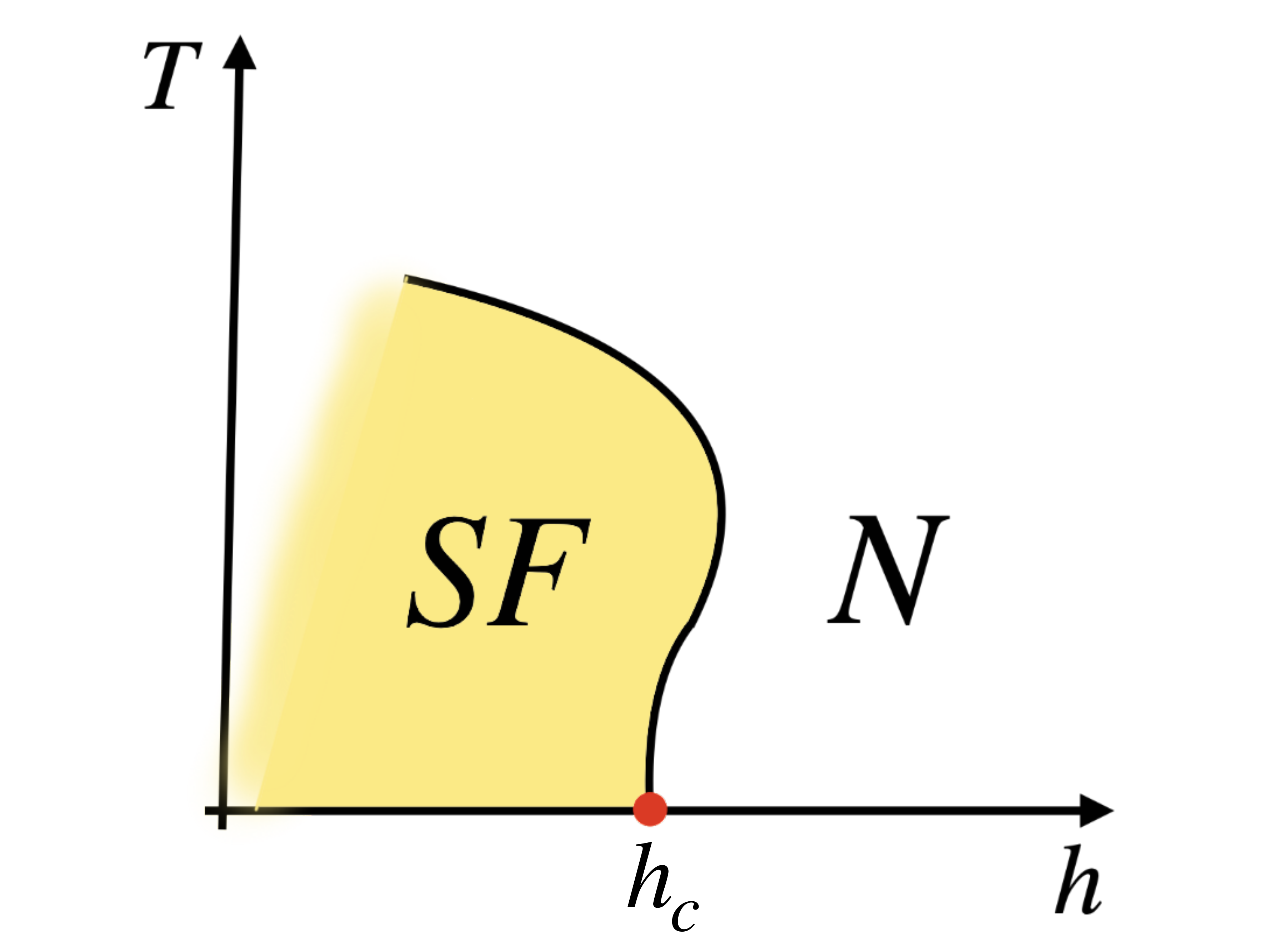}
\caption{(Color online) A schematic illustration of the mean-field phase diagram (compare Ref. \onlinecite{Zdybel_2018}). The shaded area corresponds to the superfluid phase (SF). The red dot marks the quantum phase transition located at $h=h_c$. }
\label{PD}
\end{center} 
\end{figure}

%%%%%%%%%%%%%%%%%%%%%%%%%%%%%%%%%%%%%%%%%
We now set out to analyze the structure of the pair fluctuation propagator in the normal phase. We focus on dimensionality $d=3$ and the situation, where (at mean-field level) the phase transition is continuous down to $T=0$. As was shown in Ref.~\onlinecite{Zdybel_2018}, this is achievable for the  mass ratio $r=\frac{m^-}{m^+}>r_c\approx 3$. Note that (at mean field level) this is very different from the case $d=2$, where the corresponding quantum phase transition is always discontinuous, but may be driven second-order by order-parameter fluctuation effects.\cite{Strack_2014} We treat the  "Zeeman" field $h=\frac{\mu^+-\mu^-}{2}$ as the control parameter tuning the system towards the phase transition. The system remains disordered for $h>h_c$, where $h_c$ may be expressed via the microscopic model parameters.\cite{Zdybel_2018} The occurrence of an instability towards a non-uniform (FFLO) superfluid manifests itself by a negative gradient coefficient $Z$ (see below), which is of our particular interest in Sec.~IIIA. We present a schematic mean-field phase diagram in Fig. \ref{PD}. Note that the typical profile of the phase boundary shows reentrant behavior for $r$ large enough.\cite{Zdybel_2018, Parish_2007}

We employ the Noziri{\`e}s-Schmitt-Rink method\cite{Nozires_1985} focusing on zero temperature and the vicinity of the phase transition.\cite{Liu_2006} We compute the inverse pair fluctuation propagator $F_0^{-1}(q)$ by evaluating the particle-particle bubble diagram\cite{Strinati_2018}
\begin{align}
F^{-1}_0(q)&=-\frac{1}{g}-\int_k G_{0,+}(k+q)G_{0,-}(-k)=\nonumber\\&=-\frac{m}{4\pi a_F} +\int_{\vec{k}}\left[\frac{\vartheta_{kq}^T}{iq_0-\xi_{kq}}+\frac{m}{\vec{k}^2}\right],
\label{Eq14}
\end{align}
where $m=\frac{2r}{1+r}m_+$ is the reduced mass, $\xi_{kq}=\xi_k^-+\xi_{k+q}^+$ and $\vartheta_{kq}^T=1-f(\xi_k^-)-f(\xi_{k+q}^+)$. In the above expression, we regularized the integral using the Lipmann-Schwinger equation\cite{Stoof_book}
\beq
\frac{1}{g}=\frac{m}{4\pi a_F}-\int_{\vec{k}}\frac{1}{2\varepsilon_k}\;,
\label{Eq15}
\eeq
where $a_F$ is the fermionic $s$-wave scattering length, $\varepsilon_k=\vec{k}^2/2m$ and the integrals are cutoff at momentum $\Lambda_{UV}$. The procedure eliminates ultraviolet divergencies in the limit $g\to 0^-$ and $\Lambda_{UV} \to \infty$ while keeping $a_F$ constant.\cite{Strinati_2018}

We perform analytical continuation ($iq_0\mapsto \omega+i0^+$) of  Eq.~(\ref{Eq14}), and 
 separate the obtained retarded counterpart of $F_{0}^{-1}(q)$ into real and imaginary parts using the Sokhotski formula $\frac{1}{x\pm i0^{+}}=\mathcal{P}\frac{1}{x}\mp i\pi \delta(x),$ where $\mathcal{P}$ is the Cauchy principal value. As a result, we find
\beq
\mathrm{Re}F_{0,R}^{-1}(\omega, \vec{q})=-\frac{m}{4\pi a_F} +\int_{\vec{k}}\left[\mathcal{P}\frac{\vartheta_{kq}^T}{\omega-\xi_{kq}}+\frac{m}{\vec{k}^2}\right]
\label{Eq16}
\eeq
and
\beq
\mathrm{Im}F_{0,R}^{-1}(\omega, \vec{q})=-\pi\int_{\vec{k}}\vartheta_{kq}^T\,\delta(\xi_{kq}-\omega)\;.
\label{Eq17}
\eeq
The present treatment is in a way parallel to that employed in the context of the random phase approximation for plasmons in the homogeneous electron gas,\cite{Jishi_book} where however the bosonic propagator contains the particle-hole bubble diagram instead of the particle-particle one.

\subsection{Gradient expansion and the quantum Lifshitz point}
\begin{figure}[ht] 
\begin{center}
\includegraphics[width=7.5cm]{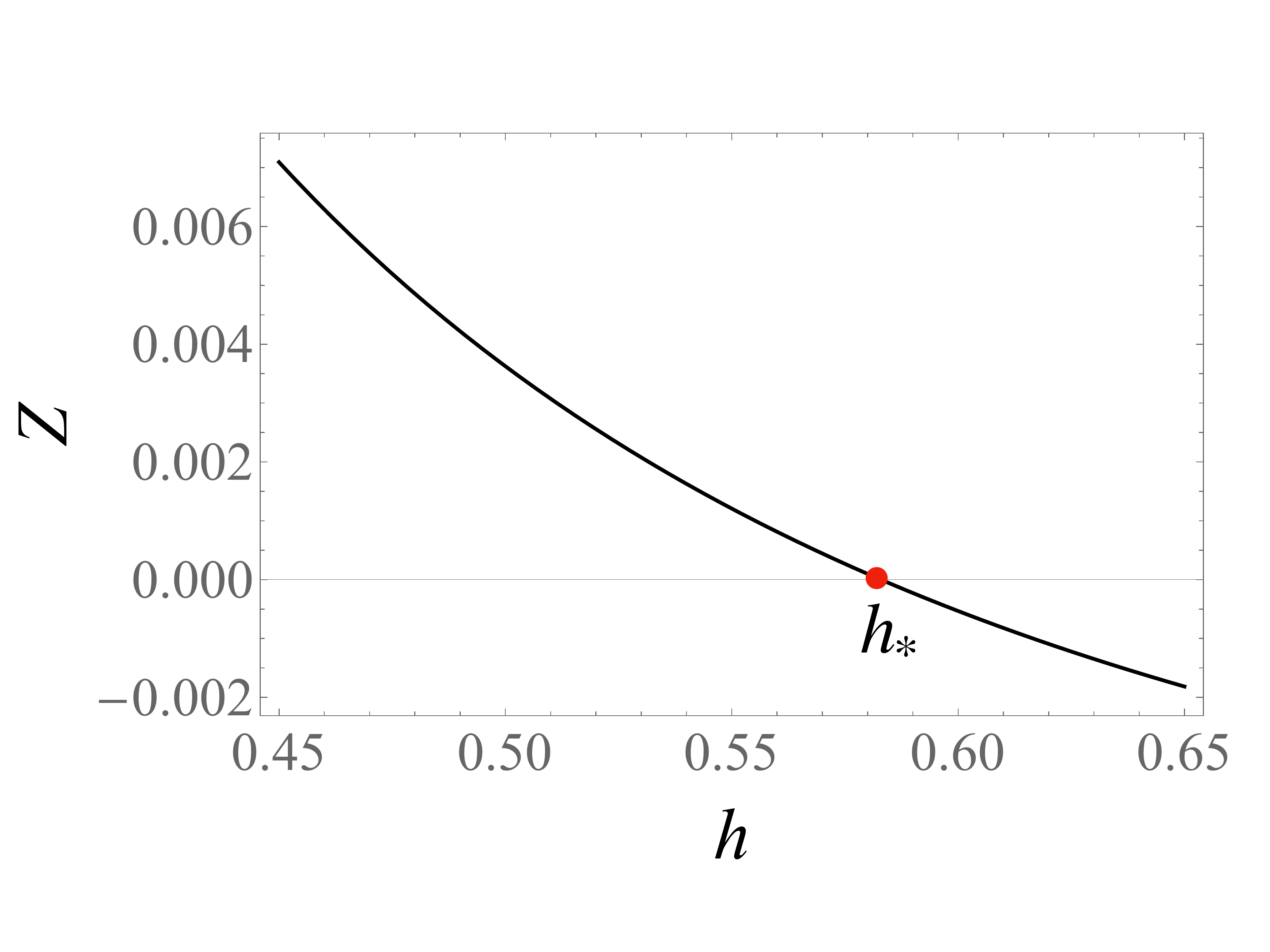}
\caption{(Color online) The gradient coefficient $Z$ as a function of the "Zeeman" field $h$. The red dot corresponds to $h_*$, where $Z=0$.  The plot parameters (see the main text) are $m^+ = 1$, $T=0$, $r = 4.03$, and $\mu = 0.1$.}
\label{Zplot}
\includegraphics[width=7.5cm]{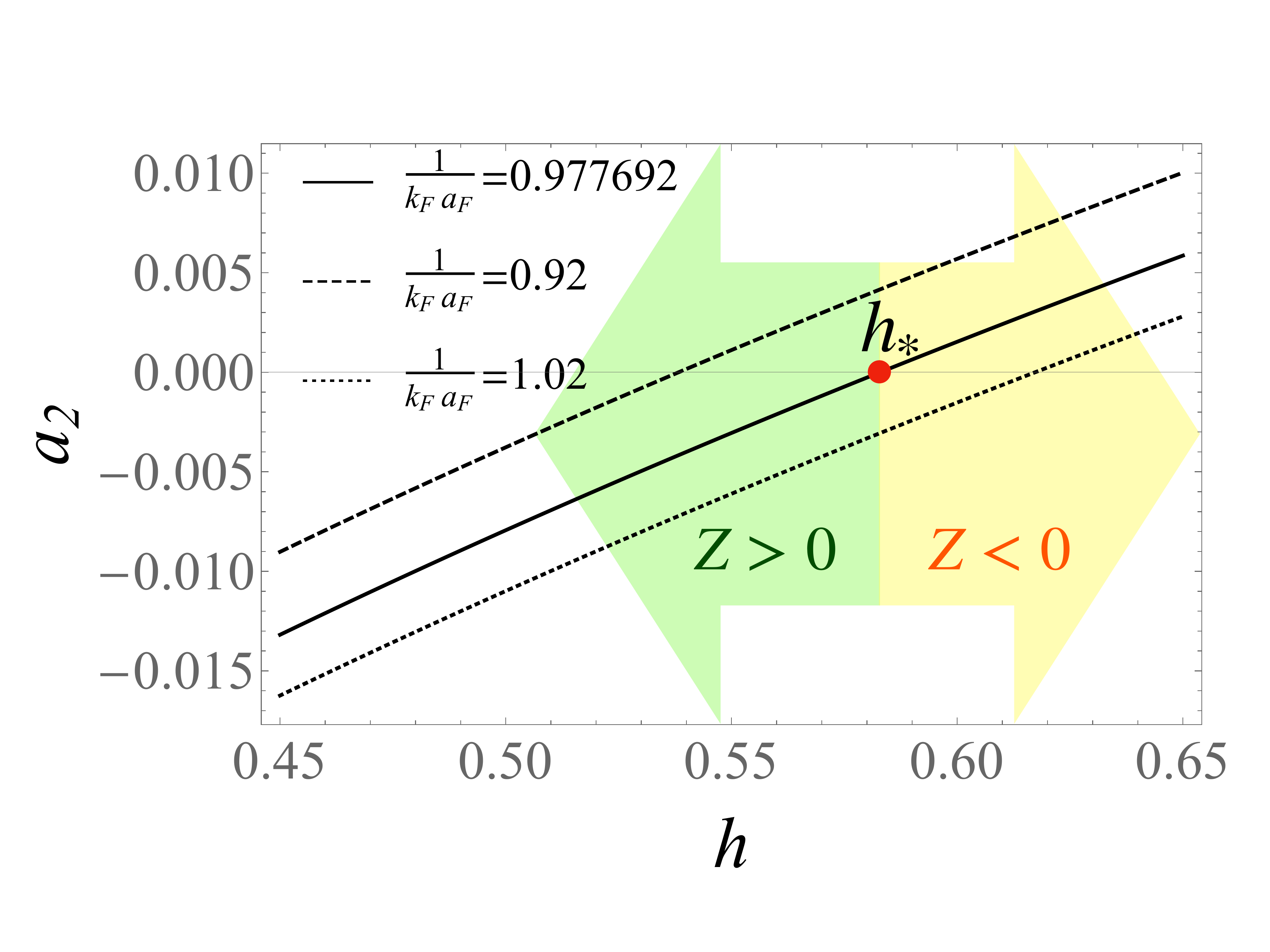}
\caption{(Color online) The quadratic Landau coefficient $a_2$ as a function of the "Zeeman" field $h$ and three values of $(k_F a_F)^{-1}$. The condition $a_2(h_c)=0$ determines the critical value $h_c$ of $h$.  The red dot corresponds to $h_*$, where $Z=0$. The quantum Lifshitz point occurs for $h_c=h_*$ and is found for $(k_F a_F)^{-1}_*=0.977692$. Notice that when $h_c>h_*$  instability towards the FFLO pairing is observed. The plot parameters (see the main text) are $m^+ = 1$, $T=0$, $r = 4.03$ ,and $\mu = 0.1$.}
\label{a2plot}
\end{center} 
\end{figure}

Focusing on $T=0$ we perform the low-frequency and low-momentum expansion of $\mathrm{Re}F_{0,R}^{-1}(\omega, \vec{q})$. We take into account terms up to second order in $\omega$ and $\vec{q}$. This leads to
\beq
\mathrm{Re}F_{0,R}^{-1}(\omega, \vec{q})=a_2+Z\vec{q}^2-W\omega-Z_0\omega^2+\dots
\label{Eq18}
\eeq
%Notice that this time, in opposition to Sec. II, we use parameterization in which we also consider a constant term in $F_{0,R}^{-1}(\omega, \vec{q})$ and we do not move it to the potential. 
The gradient coefficients $\{Z, Z_0, W\}$ as well as $a_2$ can be determined analytically in the limit $T\to0$, where $f(x)\to\theta(-x)$. 
%An analogous type of integrals, which we deal with in this case, arises when we derive the Landau expansion of the effective potential $U(\phi)$.\cite{Zdybel_2018} I
The constant contribution $\mathrm{Re}F^{-1}_{0,R}(0,\vec{0})=a_2$ corresponds to the Landau coefficient of $|\phi|^2$ and is given by\cite{Zdybel_2018}
\beq
a_2=-\frac{m}{4\pi a_F}+\frac{m}{2\pi^2}\left[k_{F,+}+\frac{k_F}{2}\ln\left(\frac{|k_{F,+}-k_F|}{|k_{F,+}+k_F|}\right)\right]\;,
\label{Eq19}
\eeq
where $k_F=\sqrt{2m\mu}$ is the "average" Fermi momentum, $k_{F,+}=\sqrt{2m^+(\mu+h)}$ denotes the "$\uparrow$"-particle ($\sigma=+$) Fermi momentum and $\mu=\frac{\mu^++\mu^-}{2}$ is the "average" chemical potential. Within mean-field theory $a_2>0$ in the normal phase, and  $a_2(h=h_c)=0$. The gradient coefficients are given by
\beq
W=\frac{m^2}{4\pi^2}\left[\frac{k_{F,+}}{k_{F,+}^2-k_F^2}-\frac{1}{2k_F}\ln\left(\frac{|k_{F,+}-k_F|}{|k_{F,+}+k_F|}\right)\right]\;,
\label{Eq20}
\eeq
\begin{align}
Z&=\frac{1}{2\pi^2}\Bigg\{\frac{m}{4}\Bigg[\frac{m}{m^+}\left(\frac{k_{F,+}}{k_{F,+}^2-k_{F}^2}-\frac{1}{2k_F}\ln\left(\frac{|k_{F,+}-k_F|}{|k_{F,+}+k_F|}\right)\right)+\nonumber\\
&+\frac{m^2}{6{m^+}^2}\left(-\frac{5k_{F,+}^3-3k_F^2k_{F,+}}{(k_{F,+}^2-k_{F}^2)^2}+\frac{3}{2k_F}\ln\left(\frac{|k_{F,+}-k_F|}{|k_{F,+}+k_F|}\right)\right)\Bigg]+\nonumber\\
&\;\;\;\;\;+\frac{m^+k_{F,+}}{(k_{F,+}^2-k_{F}^2)^2}\left[\frac{m^2k_{F,+}^2}{3{m^+}^2}-\frac{m}{2m^+}\left(k_{F,+}^2-k_F^2\right)\right]+\nonumber\\
&\;\;\;\;\;\;\;\;\;\;\;\;\;\;\;\;\;\;\;\;-\frac{mk_{F,+}\left(k_{F,+}^2-k_F^2\right)}{6 (k_{F,+}^2-k_{F}^2)^2}\Bigg\}\;,
\label{Eq21}
\end{align}
and
\beq
Z_0=\frac{m}{16\pi^2 k_F^2}\left[\frac{k^3_{F,+}+k_F^2k_{F,+}}{(k_{F,+}^2-k_F^2)^2}+\frac{1}{2k_F}\ln\left(\frac{|k_{F,+}-k_F|}{|k_{F,+}+k_F|}\right)\right]\;.
\label{Eq22}
\eeq
 The sign of $Z$ determines whether, for vanishing $a_2$, the system tends to condense in the uniform (BCS-type) or non-uniform (FFLO) ground state. We are now interested in a situation, where \emph{both} $a_2$ and $Z$ are zero, which provides a criterion for the occurrence of a (quantum) Lifshitz point. We introduce $h_*$ as a value of $h$ defined by the condition $Z(h_*)=0$ and analyze the situation, where $h_*=h_c$. It is worth noting that $\{Z, Z_0, W\}$ do not depend on the scattering length $a_F$. One may therefore tune the system to $h_*$, and independently vary $a_F$ to adjust $h_c$ towards $h_*$. In Fig.~\ref{Zplot}. we plot $Z$ for the experimentally realized mixture of $^{161}$Dy and $^{40}$K atoms,\cite{Ravensbergen_2018, Ravensbergen_2020} which is characterized by $r=4.03$ and $\mu=0.1$.
We adapt the value of $h_c$ (see Eq. \ref{Eq19}) to $h_*~$ by manipulating the parameter $(k_F a_F )^{-1}$ [see Fig. 3]. The quantum Lifshitz point corresponds to  $(k_F a_F)^{-1}_*=0.977692$ in this situation. We also note that for the physically relevant range of parameters the coefficients $W$ and $Z_0$ are positive, monotonously decreasing functions of $h$.

An analogous reasoning may be carried out for other experimentally studied Fermi mixtures characterized by sufficiently large mass imbalance $r=\frac{m^-}{m^+}>r_c\approx 3$. Examples are the $^6$Li-$^{40}$K mixture\cite{Wille_2008, Tiecke_2010, Trenkwalder_2011, Jag_2014} ($r=6.67$) as well as the $^{6}$Li-$^{53}$Cr mixture\cite{Neri_2020} ($r=8.83$). The quantum Lifshitz point is then obtained for $(k_F a_F)^{-1}_*=0.879942$ and $(k_F a_F)^{-1}_*=0.836476$ respectively (with $\mu=0.1$ in each case). In general, the value of $h_*$ is reduced upon increasing $r$ since a larger mass difference results in the occurrence of the FFLO pairing for smaller population asymmetry.

The occurrence of the quantum Lifshitz point has profound influence on the scaling behavior of the system for $T\to 0$. The RG fixed point controlling the phase transition at $T=0$ may then be non-gaussian even for $d=3$, unlike the typical situations recognized in electronic systems. Bringing the system to the proximity of the scenario discussed above should also yield complex and interesting crossover behavior.   
Resolution of the critical singular behavior at $T\to 0$ for the discussed Lifshitz transition is beyond the scope of the present paper, but certainly depends on the form of the damping term in the propagator (see the next sections), and, therefore will not fall into any of the classes discussed in the context of magnetic instabilities in electronic systems.\cite{Ramazashvili_1999, Continentino_2004} It may also be sensitive to the instability discussed in Sec.~IV. 

We finally point out, that, in a beyond-mean-field picture, the FFLO-type phases are suspected\cite{Radzihovsky_2011, Jakubczyk_2017, Wang_2018} to be marginally unstable to fluctuation effects in dimensionality $d=3$ at $T>0$ (but not at $T=0$). This would imply that, at least for infinite and homogeneous systems, the quantum Lifshitz point should \emph{always} occur in the phase diagram if there is a phase transition between the FFLO-type and uniform superfluid phases at $T=0$.

\subsection{Landau damping}
We now analyze the imaginary part of the retarded inverse propagator $\mathrm{Im}F_{0,R}^{-1}(\omega, \vec{q})$ at $T=0$.  Whenever $\mathrm{Im}F^{-1}_{0,R}(\omega, \vec{q})\neq0$,  the complex pole of $F_{0,R}(\omega, \vec{q})$ develops, which leads to broadening of the peak corresponding to a collective mode in the spectral density function $S_0(\omega, \vec{q})=-2\mathrm{Im}F_{0,R}(\omega, \vec{q})$.\cite{Altland_book, Jishi_book}  As a result, Landau damping emerges and the lifetime of collective excitations is finite. This issue was studied within the superfluid state,\cite{Kurkjian_2017, Klimin_2019, Zdybel_2019} but not in the normal phase.

We consider the limit $T\to0$ in Eq.~(\ref{Eq17}) and  introduce $u=|\vec{k}|^2>0$. Angular integration leads to
\begin{align}
\mathrm{Im}F_{0,R}^{-1}&=-\frac{m^+}{8\pi |\vec{q}|}\int_0^\infty\mathrm{d}u\left[1-\theta(-\xi_u^-)-\theta(\xi_u^--\omega)\right]\times \nonumber\\
&\times\theta\left[1-\frac{{m^+}^2}{u|\vec{q}|^2}\left(\omega-\frac{u}{m}+2\mu-\frac{\vec{q}^2}{2m^+}\right)^2\right]=
\label{Eq23} \\
&=-\frac{m^+}{8\pi|\vec{q}|}\int_0^{\lambda_\omega}\mathrm{d}u\;\theta\left[1-\frac{{m^+}^2}{u|\vec{q}|^2}\left(\omega-\frac{u}{m}+2\mu-\frac{\vec{q}^2}{2m^+}\right)^2\right],\nonumber
\end{align}
where $\lambda_\omega=2m^+r(\mu-h+\omega)$. In the above expression we observe that $\mathrm{Im}F^{-1}_{0,R}(\omega, \vec{q})$ may be nonzero only if $\omega>h-\mu$. Thus, for  values of $h=h_c>\mu$,\cite{Zdybel_2018} where the phase transition occurs, damping is absent for low energies in the normal phase.

The integrand in Eq.~(\ref{Eq23}) is equal unity in the interval $[u_-, u_+]$ and zero otherwise. Only $u_\pm >0$ are physically relevant.  These values are obtained as:
\beq
u_\pm=\frac{m^2}{2}\left[\left(A+\frac{2B}{m}\right)\pm\sqrt{\left(A+\frac{2B}{m}\right)^2-\left(\frac{2B}{m}\right)^2}\right],
\label{Eq24}
\eeq
where $A=\left(\frac{|\vec{q}|}{m^+}\right)^2$ and $B=\omega+2\mu-\frac{\vec{q}^2}{2m^+}$. The above expression is real when $\omega\geq\frac{1}{r+1}\frac{\vec{q}^2}{2m^+}-2\mu$. As we find, $\mathrm{Im}F^{-1}_{0,R}(\omega, \vec{q})\neq0$ iff $[u_-,u_+]\cap[0,\lambda_\omega]\neq\emptyset$, which divides the $(|\vec{q}|,\omega)$-plane into three distinct regions. 
\begin{figure}[ht] 
\begin{center}
\includegraphics[width=7cm]{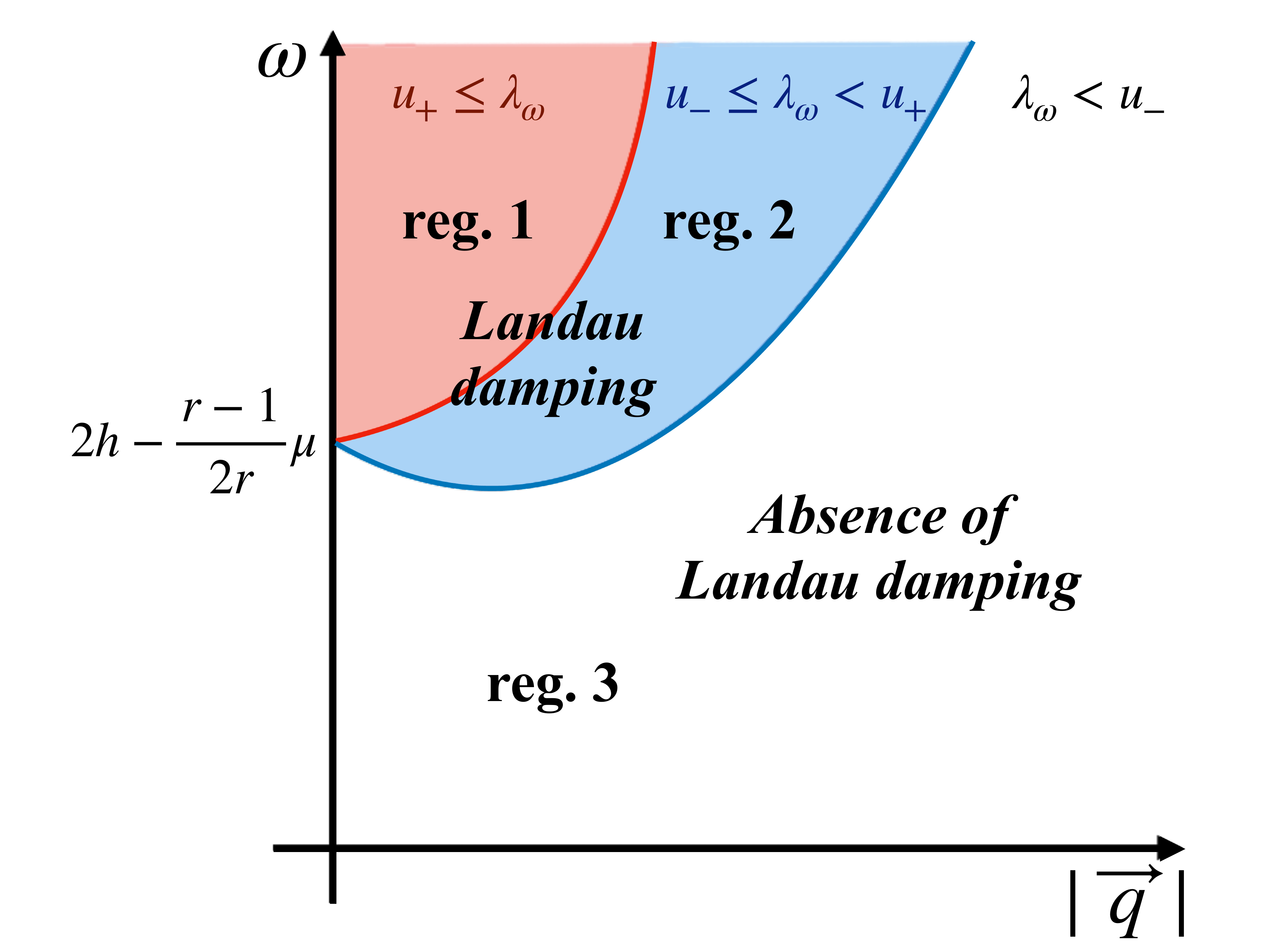}
\caption{(Color online) A schematic illustration of the $(|\vec{q}|,\omega)$-diagram at $T=0$ showing distinct regions for which Landau damping is active or absent in the normal phase.}
\label{LD}
\end{center} 
\end{figure} 
The first of them (reg.~1) is defined by $u_+\leq\lambda_\omega$. Here Landau damping is present and $\mathrm{Im}F^{-1}_{0,R}(\omega, \vec{q})$ is given by
\begin{align}
&\mathrm{Im}F_{0,R}^{-1}\left((\omega, \vec{q})\in\mathrm{reg.}~1\right)=-\frac{m^{3/2}}{4\pi}\sqrt{\omega+2\mu-\frac{1}{r+1}\frac{\vec{q}^2}{2m^+}}.
\label{Eq27}
\end{align}
Analogously, the second regime (reg.~2) is obtained from the condition $u_-\leq\lambda_\omega<u_+$. In this situation damping is active as well, but the form of  $\mathrm{Im}F^{-1}_{0,R}(\omega, \vec{q})$ reads 
\begin{align}
&\mathrm{Im}F_{0,R}^{-1}\left((\omega, \vec{q})\in\mathrm{reg.}~2\right)=
-\frac{m^+}{8\pi|\vec{q}|}\Bigg[2m^+r(\mu-h +\omega)+\nonumber\\
&-m\left\{\zeta\frac{\vec{q}^2}{2m^+}+\omega+2\mu-\frac{|\vec{q}|}{m^+}\sqrt{m\left(\omega+2\mu-\frac{1}{r+1}\frac{\vec{q}^2}{2m^+}\right)}\right\}\Bigg],
\label{Eq26}
\end{align}
where $\zeta=\frac{r-1}{r+1}$. The last regime (reg.~3) occurs whenever $\lambda_\omega<u_-$. Here $\mathrm{Im}F_{0,R}^{-1}\left((\omega, \vec{q})\in\mathrm{reg.}~3\right)=0$ and Landau damping is absent.  The obtained results are schematically summarized in the diagram of Fig. \ref{LD}.  

The obtained picture has a transparent physical interpretation. In our reasoning we identified two conditions: (a) $\omega>h-\mu$ and  (b) $\omega\geq\frac{1}{r+1}\frac{\vec{q}^2}{2m^+}-2\mu$. Violation of any of them implies that $\mathrm{Im}F_{0,R}^{-1}(\omega, \vec{q})=0$.  Recall that the quantum phase transition takes place between a fully-polarized gas and the superfluid phase. Thus,  "$\downarrow$"-particles are absent in the normal phase at $T=0$, which is implied by $\mu^-<0$ for $h>h_c>\mu$.  We may reformulate the condition (a) as $\omega>|\mu^-|>0$, which means that damping may occur only when the energy is high enough to introduce the "$\downarrow$"-particle into the system. On the other hand, we may also restate the condition (b) as $\omega\geq\xi_{pair}(\vec{q})=\frac{\vec{q}^2}{2m_{pair}}-\mu_{pair}$, where $m_{pair}=m^++m^-=m^+(1+r)$ is the mass of a pair composed of "$\uparrow$" and "$\downarrow$" particles, and $\mu_{pair}=\mu^++\mu^-=2\mu$ is its chemical potential. Therefore, damping may be active only when the excitation of a pair carrying momentum $\vec{q}$ is achievable. 

However, conditions (a) and (b) do not take into account the momentum conservation, which leads to further restrictions of areas in which  $\mathrm{Im}F_{0,R}^{-1}(\omega, \vec{q})\neq 0$. In particular, the step function in Eq. (\ref{Eq23}) inflicts values of $\omega$ and $\vec{q}$, which are consistent with possible directions of the momentum given by $\cos \theta=\vec{k}\cdot\vec{q}/|\vec{k}||\vec{q}|\in[-1,1]$.

We note that $\mathrm{Im}F_{0,R}^{-1}(\omega, \vec{q})$ is a non-positive function of $\omega$ and $|\vec{q}|$ in both regions where Landau damping is active. This feature ensures that the spectral density function $S_0(\omega, \vec{q})$ is non-negative for all values of $\omega$ and $|\vec{q}|$. %Thus, obtained here expressions agree with constraints impose on $S_0(\omega, \vec{q})$.\cite{Nolting_book}

Notice, that at $|\vec{q}|$ small damping occurs for $\omega>2h-\frac{r-1}{2r}\mu>0$, because  $h_c>\mu$.\cite{Zdybel_2018} Therefore, in the proximity of the quantum phase transition within the normal phase, the damping terms do not appear in the low-frequency and low-momentum expansion. As already emphasised,  this situation is entirely different as compared to the symmetry-broken phase (see Sec.~II). 

In the following Sec.~IV we analyze the impact of the Landau damping in the form discussed in Sec.~II on the renormalization group flow in the vicinity of the quantum critical point.

%%%%%%%%%%%%%%%%%%%%%%%%%%%%%%%%%%%%%%%%%
\section{Renormalization group flow} 
%%%%%%%%%%%%%%%%%%%%%%%%%%%%%%%%%%%%%%%%%
We now focus on the situation, where the phase transition between the normal and uniform superfluid phases is second order (and not multicritical). We examine fluctuation effects within the renormalization group framework. We connect to the Hertz-Millis theory\cite{Hertz_1974, Millis_1993}  which originally was formulated with reference to the symmetric phase and addressed magnetic phase transitions in itinerant Fermi systems. %From this point of view, the picture obtained at the mean-field level is clearly incomplete. For example the correct universal asymptotic behavior of the critical line can only be recovered taking fluctuation effects into account. 
An important insight of the Hertz-Millis theory is that the (second-order) quantum phase transition at $T=0$ belongs to the universality class of the appropriate $O(N)$-symmetric model in effective dimensionality $D=d+z$, where the dynamical exponent $z$ takes the value $z=3$ for instabilities at ordering wavevector $\vec{Q}=0$. The value of $z$ stems from Landau damping and %; in its absence we would expect $z=1$.\cite{Sachdev_book}
the present situation is, from this perspective, not immediately obvious, since Landau damping is present only in the symmetry-broken phase and the coefficient of the term $\sim\frac{|q_0|}{|\vec{q}|}$ is proportional to $\rho_0$ and therefore vanishes at the critical point. We further note that the quantum critical point corresponds to a state, which is ordered at the level of the mean-field theory so that the Landau-damping term is present in the entire (critical) renormalization group flow and vanishes only for asymptotically low RG scales. The Hertz-Millis theory is, on the other hand, formulated in the symmetric phase. %A direct implementation of this framework to the presently studied system, ignoring the properties of the low-$T$ phase, would yield a quantum critical point with $z=1$ (very similar to the case of the quantum Ising model in $d=3$). 
%This treatment is however not fully justified. 
Below we employ a non-perturbative renormalization group framework applicable to both symmetric and symmetry-broken phases\cite{Jakubczyk_2008} and capturing both the Gaussian and Wilson-Fisher fixed points. We show how the expected quantum-classical crossover at $T>0$ is accurately recovered by our approach in absence of Landau damping. We subsequently include the Landau damping term $\sim \rho_0 \frac{|q_0|}{|\vec{q}|}$. Our results indicate that this leads to an obstruction of the RG flow towards the Wilson-Fisher fixed point occurring at intermediate RG scales and point towards a possible occurrence of a fluctuation-induced weakly first order phase transition driven by Landau damping of the longitudinal ($\sigma$) mode. The mechanism of destabilizing the quantum critical point active here is entirely different from those studied before\cite{Belitz_2005, Lohneysen_2007} e.g. for the ferromagnetic transitions\cite{Belitz_2005_2} or superconductors.\cite{Halperin_1974, Li_2009}    

We employ the one-particle-irreducible variant of nonperturbative renormalization theory, taking the Wetterich equation\cite{Wetterich_1993} 
\beq
\label{Wetterich_eq}
\dot{\Gamma}=\frac{1}{2}\int_q\left(\dot{R}^\sigma G_\sigma + \dot{R}^\pi G_\pi  \right)\;
\eeq 
as the starting point. This framework is exceptionally convenient for resolution of crossover phenomena due to multiple fixed points governing the RG flow at distinct scales (see e.g. Refs.~\onlinecite{Strack_2009,  Leonard_2015, Lammers_2016, Debelhoir_2016, Rancon_2017, Chlebicki_2019}). Eq.~(\ref{Wetterich_eq}) describes the renormalization group flow of the effective action $\Gamma =\Gamma_t[\phi]$ upon reducing the RG scale, implemented here as the momentum cutoff $\Lambda=\Lambda_0e^t$ with $t\in (-\infty,0)$ and $\Lambda_0$ denoting the microscopic momentum scale. The quantity  $\Gamma =\Gamma_t[\phi]$ thus continuously connects the microscopic action $\mathcal{S}[\phi]=\Gamma_{t\to 0}[\phi]$ and the thermodynamic free energy $F[\phi]=\Gamma_{t\to-\infty}[\phi]$.

 In the shorthand notation used in Eq.~(\ref{Wetterich_eq}) the dot indicates differentiation with respect to the logarithmic scale $t=\log\left(\Lambda/\Lambda_0\right)$, i.e. $\dot{X}=\partial_tX$, the two-component real field $\phi$ is decomposed (compare Sec.~II) into the longitudinal and transverse modes $\phi(x)=\phi_0+\sigma(x) + i\pi(x)$ with $\phi_0$  chosen real, ${R}^{\sigma/\pi}$ denotes the cutoff function added to the inverse $\sigma/\pi$ propagator to regularize the low-momentum fluctuations. Finally, $G_{\sigma/\pi}$ is the regularized $\sigma/\pi$ propagator, obtained by taking the second field derivative of the regularized action in the $\sigma/\pi$ direction. For a detailed exposition of the framework, we refer to Refs.~\onlinecite{Berges_2002, Kopietz_book, RG_book, Dupuis_2020}. The approximation strategy taken below is  an adaptation of the framework developed in previous studies of magnetic phase transitions in Fermi systems with Ising symmetry\cite{Jakubczyk_2008, Jakubczyk_2009_phi6, Jakubczyk_2010} and conventional Hertz-Millis action. For the reason of highly anisotropic propagator obtained in the previous sections [Sec.~II and III], we use two different cutoff functions $R^{\sigma/\pi}$ for the distinct directions. 
 %We are here mostly interested in the effect of the Landau damping term in the $\sigma$ propagator on the RG flow in the symmetry-broken phase. 
 We aim here at distilling the impact of the specific form of the Landau-damping term on the quantum-critical scaling, and, for the sake of simplicity, we will  neglect the off-diagonal $\sigma-\pi$ propagator. This term is absent for particle-hole symmetric systems and is supposed to scale to zero under RG flow in the entire symmetry-broken phase.\cite{Obert_2013} In generality, it however influences  some of the transport and thermodynamic properties, such as the condensate compressibility.\cite{Pistolesi_2004}
 
  With this simplified picture in mind, proceeding along the standard track,\cite{Berges_2002} we obtain the flow of the effective potential $U(\rho)$ by evaluating the Wetterich equation [Eq.~(\ref{Wetterich_eq})] at a constant field configuration. This yields   
\beq
\label{U_flow}
\dot{U}(\rho)=\frac{1}{2}\int_q\left\{\dot{R}^\sigma [\gamma_\sigma (\rho)]^{-1}  +  \dot{R}^\pi [\gamma_\pi (\rho)]^{-1} \right\}\;,
\eeq 
where, within our parametrization, the inverse propagators are  given by: 
\beq
\label{gamma_sigma}
\gamma_\sigma(\rho)= Z_\sigma \vec{q}^2+Z_\sigma^0q_0^2+L\rho_0\frac{|q_0|}{|\vec{q}|}+U'(\rho)+2\rho U''(\rho)+R^\sigma
\eeq 
and 
\beq
\label{gamma_pi}
\gamma_\pi(\rho)= Z_\pi \vec{q}^2+Z_\pi^0q_0^2+U'(\rho)+R^\pi.
\eeq 
%We neglect the dependence of the gradient coefficients on $\rho$. 
In what follows, we parametrize the flowing effective potential with the simple quartic form
\beq
U(\rho)=\frac{\lambda}{2}\left(\rho-\rho_0\right)^2\;.
\eeq
In the present approximation, we also neglect the flow of the gradient coefficients (both in the space and time directions). This is equivalent to dropping the anomalous dimensions (which are anyway small for the three-dimensional systems studied here). In consequence, the RG flow is parametrized by a set of only two flowing couplings ($\rho_0$ and $\lambda$). In  contexts concerning quantum criticality in itinerant Fermi systems (see. e.g. Ref.~\onlinecite{Jakubczyk_2010}) this simple approximation level captures the relevant aspects of physics in dimensionality $d=3$. Note however, that it is quite insufficient in $d=2$, where the flow of the $Z$-factors and the related anomalous dimensions play a prominent role.
% The important observation here is that the Landau damping term in $\gamma_\sigma$ [Eq.~(\ref{gamma_sigma})] will asymptotically scale to zero under RG if the system is at the critical point. %However, at $T>0$ the system properties are influenced by both the quantum ( high $\Lambda$) and classical (low $\Lambda$) regimes of the flow. In consequence, the impact of Landau damping on the phase diagram is not immediately clear at all. 

We now extract the flow of $\rho_0$ from $\frac{d}{dt}U'(\rho_0)=\partial_t U'(\rho_0)+U''(\rho_0)\dot{\rho_0}$. 
Differentiating Eq.~(\ref{U_flow}) we obtain
\beq 
\label{rho_dot}
\dot{\rho_0}=\frac{1}{2}\int_q\left(3\dot{R}^\sigma\gamma_\sigma^{-2}  + \dot{R}^\pi\gamma_\pi^{-2} \right)  \;, 
\eeq 
where we denoted $\gamma_{\sigma /\pi}(\rho_0)=\gamma_{\sigma/\pi}$
The flow of the quartic interaction coupling $\lambda$ is obtained by taking the second $\rho$-derivative of Eq.~(\ref{U_flow}) and evaluating at $\rho=\rho_0$. This yields: 
\beq
\label{lam_dot}
\dot{\lambda}=\lambda^2\int_q \left(9\dot{R}^\sigma\gamma_\sigma^{-3}  + \dot{R}^\pi\gamma_\pi^{-3} \right)  \;. 
\eeq
Note that since $U'(\rho_0)=0$, putting $\rho=\rho_0$ leads to a massless $\pi$-propagator $\gamma_\pi^{-1}$ [Eq.~(\ref{gamma_pi})]. The mass of the $\sigma$-propagator is on the other hand given by $m_\sigma^2=U'(\rho_0)+2\rho_0U''(\rho_0)=2\lambda\rho_0$. As concerns the cutoff: in the following calculation we implement the direction-dependent Litim cutoff\cite{Litim_2001} 
\beq
R^{\sigma / \pi} = \mathcal{X}_{\sigma / \pi}  \theta(\mathcal{X}_{\sigma / \pi} )\;,
\eeq
where  
\beq
\mathcal{X}_\sigma = Z_\sigma (\Lambda^2-\vec{q}^2)-Z_\sigma^0 q_0^2-L\rho_0\frac{|q_0|}{|\vec{q}|}
\eeq 
  and 
 \beq
 \mathcal{X}_\pi = Z_\pi (\Lambda^2-\vec{q}^2)-Z_\pi^0 q_0^2\;. 
 \eeq 
With this convenient choice the integrals in the flow equations for $\rho_0$ and $\lambda$ become effectively constrained to the regions of the $(q_0,\vec{q})$ space, where $\gamma_\sigma$ and $\gamma_\pi$ are constant (i.e. $q$-independent), which yields the momentum integrals straightforward. In what follows, we use the set of variables rescaled according to the canonical classical dimensions:
\beq 
\kappa = Z_\pi\Lambda^{2-d}\rho_0\;,\;\;\;\; u=Z_\pi^{-2}\Lambda^{d-4}\lambda\;,
\eeq   
so that at $T>0$ the flow manifestly reaches the (Wilson-Fisher) fixed point if the initial condition for the flow is chosen at the critical manifold. 
The right-hand sides of the flow equations are  split into the classical ($q_0=0$) and quantum ($q_0\neq 0$) contributions, so that 
\beq 
\label{kappa_flow}
\dot{\kappa}=\beta_{\kappa}^{cl} + \beta_{\kappa}^{q,\pi} +  \beta_{\kappa}^{q,\sigma}
\eeq  
\beq 
\label{u_flow}
\dot{u}=\beta_{u}^{cl} + \beta_{u}^{q,\pi} +  \beta_{u}^{q,\sigma}\;.
\eeq    
We additionally separated the quantum contributions from the longitudinal ($\sigma$) and transverse ($\pi$) fluctuations  [compare Eq.~(\ref{rho_dot}) and (\ref{lam_dot})]. The Matsubara summations occurring in the quantum contributions must be carried our numerically. The explicit expressions for the flow equations are given in the appendix. The flow equations are written for arbitrary dimensionality $d$, from now on we however restrict to $d=3$. As already mentioned, the present approximation level is inadequate for the case $d=2$ since the anomalous dimensions are neglected. The critical singularity is controlled by the Gaussian fixed point at $T=0$ and by the classical Wilson-Fisher fixed point for $T>0$. 

\subsection{RG flow in absence of Landau damping}
We now summarize the renormalization group flow neglecting the Landau damping of the longitudinal mode. This amounts to putting $L=0$ in  $\beta_{\kappa}^{q,\sigma}$ and $\beta_{u}^{q,\sigma}$. The flow is initiated at $s=-t=-\log(\Lambda/\Lambda_0)=0$ with fixed temperature $T>0$ and given values of $\kappa=\kappa_0>0$ and $u=u_0>0$ deriving from the microscopic values. In the numerical solutions presented below we put $u_0=\Lambda_0=Z_\sigma=Z_\pi=Z_\sigma^0=Z_\pi^0=1$.  Integrating the flow towards large positive values of $s$ (corresponding to $\Lambda\to 0$) we find either $\kappa$ reaching zero at a finite scale, indicating flow into the symmetric phase, or its divergence for $s$ large (corresponding to $\rho_0$ converging to a finite value), indicating the symmetry-broken state. By tuning the system towards the phase transition, we observe the flow of $(\kappa, \;u)$ towards fixed-point values. The results for the flow of $\kappa$ are exhibited in Fig.~5 for a sequence of temperatures approaching zero. The crossover scale, where the flow departs from the Gaussian fixed point (governing the critical singularity at $T=0$) and flows towards the Wilson-Fisher fixed point (controlling the critical singularity at $T>0$) diverges for $T\to 0^+$ according to $s_{cross}\sim -\frac{1}{z}\log (T)$ [see Ref.~(\onlinecite{Millis_1993})]. A data fit gives $z=1.0$ in full accord with the expected behavior. The momentum range $(0,s_{cross})$ corresponds to the flow regime dominated by the quantum ($q_0\neq 0$) contributions, while the range $(s_{cross}, \infty)$ is dominated by the classical ($q_0$) contribution to the flow equations. 
\begin{figure}[ht] 
\label{Fig}
\begin{center}
\includegraphics[width=8cm]{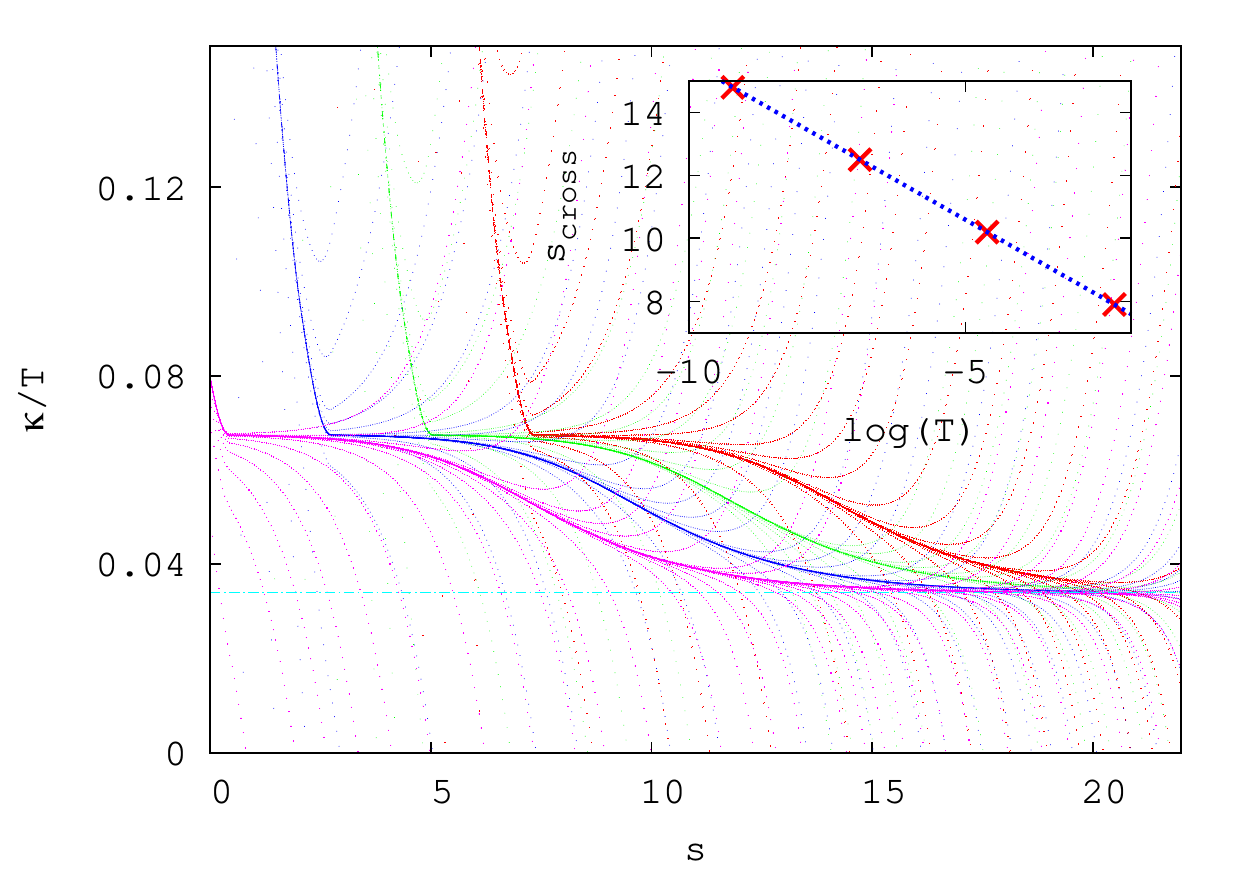}
\caption{(Color online) Renormalization flows of $\kappa$ for a sequence of temperatures: $T=10^{-4}$ (the uppermost curve), $T=10^{-3}$, $T=10^{-2}$, $T=10^{-1}$ (the lowermost curve) in the absence of Landau damping. The intermediate regime of the flow is controlled by the Gaussian fixed point, while the infrared sector ($s$ large) is governed by the Wilson-Fisher fixed point. The crossover scale $s_{cross}$, where the flow departs from the Gaussian fixed point diverges for $T\to 0$. One generally expects\cite{Millis_1993} $s_{cross}\sim -\frac{1}{z}\log (T)$. This is in full agreement with our results with $z=1.0$, as demonstrated by the inset plot.    }
\end{center} 
\end{figure} 

We now present the results obtained for the critical line $T_c(\kappa_0)$ by integrating the RG flow. The general expectation\cite{Millis_1993} yields (for $T$ low) the power-law behavior  
\beq 
\label{psi_formula}
T_c\sim \left(\kappa_0-\kappa_0^{(0)}\right)^\psi \;\;\;\textrm{with \;\;\;} \psi=\frac{z}{d+z-2}\;.
\eeq
Note that $\psi$ as given by this formula, depends exclusively on $d$ and $z$ and appears completely insensitive to the nature of the symmetry-broken phase. Exemplary results for the critical line, obtained by integrating the flow equations [Eq.~(\ref{kappa_flow}), (\ref{u_flow})] are presented in Fig.~6.
\begin{figure}[ht]
\begin{center}
\label{T_c_plot}
\includegraphics[width=8cm]{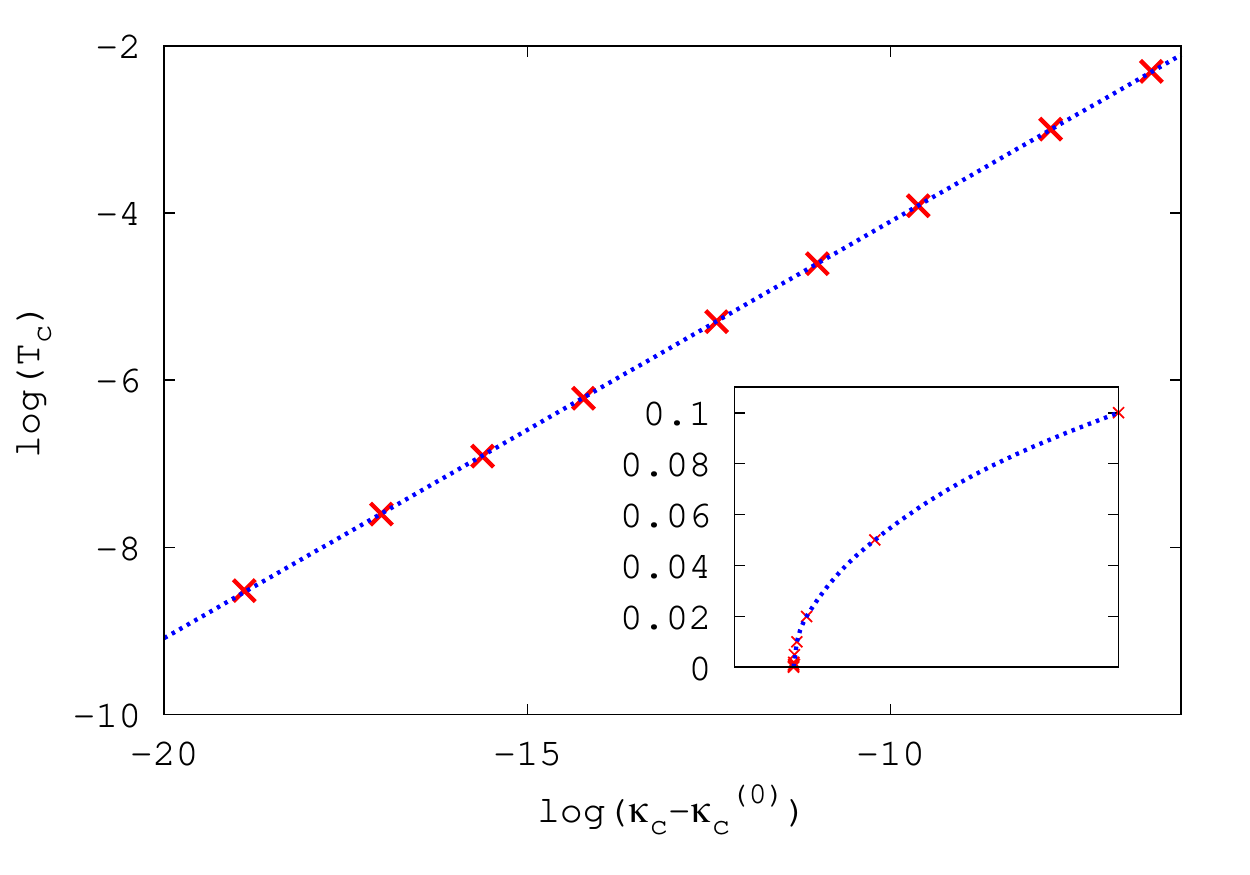}
\caption{(Color online) The critical line evaluated from integrating the RG flow in the absence of Landau damping ($L=0$). The phase boundary follows a power law with $\psi=0.5$ in full agreement with the prediction of Eq.~(\ref{psi_formula}) with $z=1$. In the inset we present the same data plotted in linear scale.}
\end{center}
\end{figure}   
The obtained power law yields $\psi=0.5$ in full agreement with the expected behavior with $z=1$ ($\psi=1/2$). Also observe, that the universal shape of the $T_c$-line is incompatible with the reentrant behavior quite generically obtained at mean field level (see the illustrative Fig.~1, and e.g. Ref.~\onlinecite{Zdybel_2018}).

The above calculation reproduces the anticipated picture of quantum criticality upon dropping the Landau damping term in the flow equations. We emphasize the role of the dynamical exponent, which governs both the $T_c$-line and the quantum-classical crossover scale. It is now our aim to investigate how this picture becomes deformed if the Landau damping contribution to the $\sigma$-mode propagator (as found in the effective action analyzed in Sec.~II and ~III) is included.   
\subsection{RG flow in presence of Landau damping}  
We proceed by repeating the analysis of the previous subsection for nonzero Landau damping coupling $L$. We solve the flow equations [Eq.~(\ref{kappa_flow}-\ref{u_flow})] following the procedure of tuning  $\kappa_0$ towards its critical value for a sequence of temperatures approaching zero. The quantum regime of the flow is now governed by the interplay of the distinct contributions (coming from the terms $\sim q_0^2$ and $\sim\rho_0\frac{|q_0|}{|\vec{q}|}$). The analysis of the flow of $\kappa$ is depicted in Fig.~7 for a relatively small value of $L$ ($=0.1$) and low temperature. The flow is superimposed with the corresponding situation obtained at the same $T$ (as well as the remaining parameters) for $L=0$. The dichotomy procedure as described in the previous subsection leads to identification of the two phases present in the system as well as the phase boundary in the parameter space. The phase transition is however not accompanied by the convergence of the flow to the fixed point and the associated scale-invariance. This behavior may indicate a fluctuation-induced first-order transition originating from Landau damping. 
 \begin{figure}[ht] 
\label{Fig}
\begin{center}
\includegraphics[width=8cm]{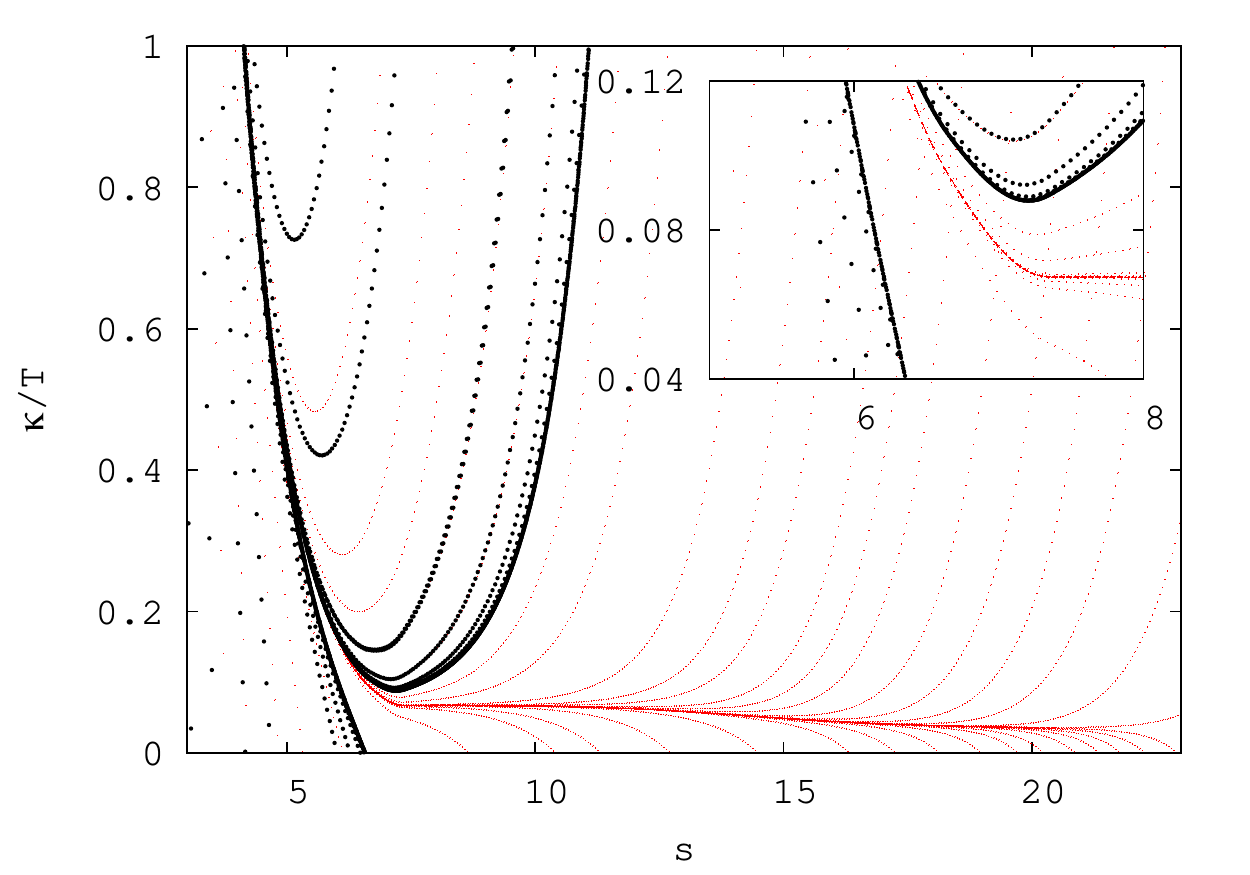}
\caption{(Color online) Renormalization flow of $\kappa$ for $T=10^{-4}$ in presence of Landau damping with $L=0.1$ (black points) superimposed with the corresponding flow for $L=0$ (light red points). The presence of Landau damping obstructs the flow towards the fixed-point. The phase transition is still present (as signaled by $\kappa$ either reaching zero, or flowing to infinity), but the system does not exhibit scale invariance. This indicates a first-order phase transition. The inset is the zoom of the region where the RG trajectories separate.   }
\end{center} 
\end{figure}  
  The effect occurs only for temperatures sufficiently low, at $T$ higher we find a continuous transition as before. This would imply the presence of a tricritical point at a temperature $T_{tri}$ in the phase diagram. We note that for $T>T_{tri}$ the system shows behavior very similar to what we obtained for $L=0$. The value of $T_{tri}$ increased with $L$ and for $T\ll T_{tri}$ the critical value of the control parameter $\kappa_0$ is practically independent of $T$, so that the $T_c$-line in the $(\kappa_0, T)$ plane is a vertical straight line. The obtained $T_c$-line is plotted in Fig.~8 for a sequence of values of $L$. Clearly visible is the deviation from the scaling behavior with $z=1$ at low $T$. 
 \begin{figure}[ht] 
\label{Fig}
\begin{center} 
 \includegraphics[width=8cm]{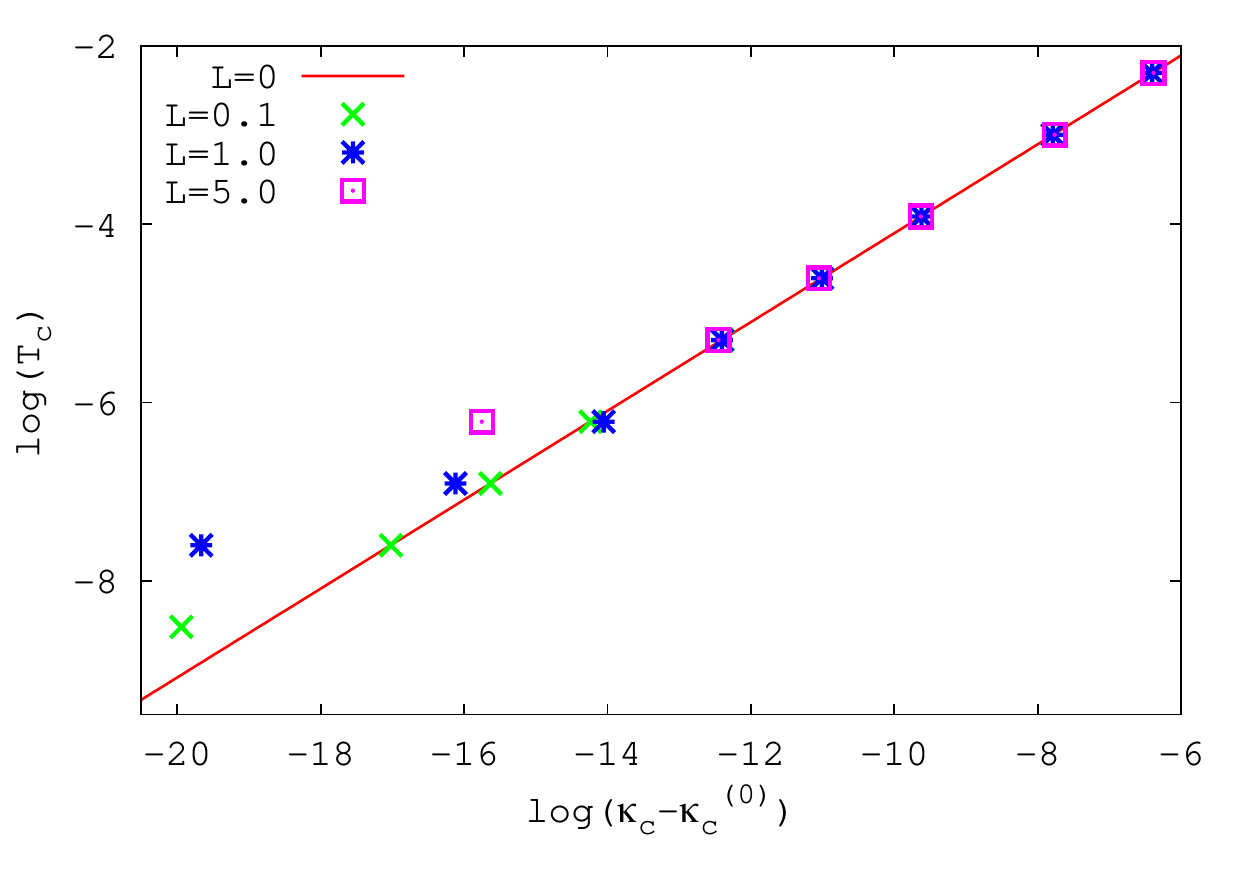}
\caption{(Color online) The critical temperature as function of $\kappa_0$ obtained for a sequence of values of $L$. At high temperatures the $T_c$-line follows the power-law behavior with the exponent $\psi=1/2$. A deviation from this scaling occurs below a threshold value of $T$ (which depends on $L$). Below the lowest-lying point in each of the plotted curves, the critical value of $\kappa_0$ is $T$-independent, so that $\log(\kappa_0-\kappa_0^{(0)})$ approaches $-\infty$.   
  }
\end{center} 
\end{figure}  
Below the lowest-lying point in each of the plotted curves, the critical value of $\kappa_0$ is $T$-independent, so that $\log(\kappa_0-\kappa_0^{(0)})$ approaches $-\infty$.  

We close this section by remarking that the obtained deviation from the quantum-critical scaling with $z=1$, signaling a possible first-order transition is visible only for very low temperatures. We obtained no scaling regime characteristic of $z=3$. We finally point out, that the damping-induced obstruction of the flow towards the Wilson-Fisher fixed point was demonstrated here only within a very simple parametrization of the flowing effective action and for a $\phi^4$-type initial condition. We can make no statement concerning the generality of this phenomenon beyond this approximation level.    

%%%%%%%%%%%%%%%%%%%%%%%%%%%%%%%%%%%%%%%%%
\section{Conclusion and outlook}
%%%%%%%%%%%%%%%%%%%%%%%%%%%%%%%%%%%%%%%%%
In this paper we addressed two distinct aspects concerning the superfluid phase transition in imbalanced Fermi mixtures. By analyzing the structure of the effective action for the order-parameter field, we have shown that, for experimentally realized mixtures and physically relevant sets of parameters, the Lifshitz point located between the normal, FFLO-type and uniform superfluid phases may be tuned to zero temperature. A resulting Lifshitz quantum critical point (or its proximity) would present in dimensionality $d=3$ a not-yet-explored situation involving a potentially non-Gaussian phase transition at temperature $T=0$. This may have important  consequences for the system, which require further studies. For example, the scaling shape of the $T_c$-line might no longer be described by Eq.~(\ref{psi_formula}), but instead be controlled by the correlation length exponent $\nu$ akin to the case of the two-dimensional quantum Ising model.\cite{Sachdev_book} Note however that there are two distinct critical exponents\cite{Diehl_2002} ($\nu_\perp$ and $\nu_\parallel$) characterizing the behavior of the correlation function at the Lifshitz point.

 %The Lifshitz point analyzed by us here is quite different from those invoked in the context of antiferromagnetic systems,\cite{Ramazashvili_1999} where $z=2$ and the corresponding phase transition at $T=0$ is Gaussian. 

We subsequently studied the pairing field propagator across the superfluid phase transition (not necessarily of Lifshitz type). We established a characteristic feature of the presently addressed system which concerns the Landau damping. This turns out to occur  exclusively on the symmetry-broken side of the phase transition and affect only the longitudinal ($\sigma$) mode. Landau damping manifests itself by the presence of a term $\sim\rho_0\frac{|q_0|}{|\vec{q}|}$ in the inverse $\sigma$-propagator. With the aim of assessing the role of this term, we performed a renormalization group calculation. Our results indicate an obstruction of the RG flow towards the scaling solution (Wilson-Fisher fixed point). Physically, this indicates a possibility of obtaining a weakly first-order transition due to the coupling between the order-parameter mode and fermionic excitations (effectively described by the damping term). The obtained instability occurs only at very low temperatures. Upon increasing $T$ a quantum-critical scaling characteristic of the dynamical exponent $z=1$ is recovered, so that Landau damping has no impact on the physical properties. In a realistic situation and away from the Lifshitz point scenario, one may therefore expect a scaling behavior resembling the $d=3$ quantum Ising model, but cutoff at low $T$. 
A similar instability may also affect the quantum Lifshitz point analyzed in the first part of the paper.       

The recently realized Fermi mixtures involving large mass imbalance constitute candidates for an experimental realization of the situation analyzed in the present paper. One promising candidate is the mixture of $^{161}$Dy and $^{40}$K atoms\cite{Ravensbergen_2018, Ravensbergen_2020}, which, due to  the existence of a broad Feshbach resonance\cite{Chin_2010} can be tuned more flexibly than the earlier studied mixtures of $^6$Li and $^{40}$K atoms.\cite{Wille_2008, Tiecke_2010, Trenkwalder_2011, Jag_2014}

\begin{acknowledgments}
We acknowledge support from the Polish National Science Center via 
2014/15/B/ST3/02212 and 2017/26/E/ST3/00211. 
\end{acknowledgments}

%%%%%%%%%%%%%%%%%%%%%%%%%%%%%%%%%%%%%%%%%
\section*{Appendix} 
%%%%%%%%%%%%%%%%%%%%%%%%%%%%%%%%%%%%%%%%%

Below we give  explicit expressions for the distinct contributions occurring in the flow equations [Eq.~(\ref{kappa_flow}-\ref{u_flow})]. For the flow of $\kappa$ we obtained: 
\beq
\beta_{\kappa}^{cl} = \left(2-d\right)\kappa + \frac{A_d T}{d}\left(\frac{3z}{(z+{\tilde{m}_\sigma}^2)^2}+1\right)\;, 
\eeq
\beq
\label{beta_pi}
\beta_{\kappa}^{q,\pi}=\frac{2A_dT}{d}\sum_{q_0>0}^{\tilde{q}_0^\pi}\left[y_\pi^M\left(q_0\right)\right]^{d/2}\;,
\eeq 
\begin{align}
\label{beta_sigma}
&\beta_{\kappa}^{q,\sigma}=\frac{3A_dT}{(z+{\tilde{m}_\sigma}^2)^2}\sum_{q_0>0}^{\tilde{q}_0^\sigma}\Bigg[\frac{2z}{d}\left(\left[y_\sigma^M\left(q_0\right)\right]^{d/2}-\left[y_\sigma^m\left(q_0\right)\right]^{d/2}\right)  \\
&-\frac{q_0LZ_\pi^{-2}\Lambda^{d-5}}{d-1}\left(\dot{\kappa}+(d-2)\kappa\right) \left(\left[y_\sigma^M(q_0)\right]^{(d-1)/2}-\left[y_\sigma^m(q_0)\right]^{(d-1)/2}\right)   \Bigg]\;,\nonumber
\end{align}
where $A_d=\frac{S^{d-1}}{(2\pi)^d}$ with $S^{d-1}$ denoting the surface area of the $(d-1)$-dimensional unit sphere, $z=Z_\sigma/Z_\pi$, $\tilde{m}_{\sigma}^2=2\kappa u$, $y_\pi^{M}(q_0)=1-\frac{Z_{\pi}^0}{Z_{\pi}\Lambda^2}q_0^2$, while $y_\sigma^{M/m}(q_0)$ are the (positive) roots of the equation
\beq 
y^{3/2}-\left(1-\frac{Z_{\sigma}^0}{Z_\sigma\Lambda^2}q_0^2\right)y^{1/2}+\frac{L\rho_0}{Z_\sigma\Lambda^3}|q_0|=0\;. 
\eeq	 
The Matsubara summation for the $\pi$ contribution [Eq.~(\ref{beta_pi})] runs from $q_0=2\pi T$ up to $\tilde{q}_0^{\pi} = \sqrt{\frac{Z_{\pi}}{Z_{\pi}^0}}\Lambda$. For the $\sigma$ contribution [Eq.~(\ref{beta_sigma})], the summation over $q_0=2 n \pi T$ ($n\in \mathbb{N}^+$) is restricted to values fulfilling the inequality
\beq
q_0^2\frac{Z_\sigma^0}{Z_\sigma \Lambda^2}+3(2)^{-2/3}\left(\frac{L\rho_0}{Z_\sigma\Lambda^3}q_0\right)^{2/3}-1<0\;.
\eeq
The largest value of $q_0$ fulfilling this condition is identified as $\tilde{q}_0^{\sigma}$. 

The corresponding contributions to the flow of the interaction coupling $u$ may be expressed as follows:
\beq
\beta_{u}^{cl} = \left(4-d\right)u +  \frac{2A_d T}{d} \left(\frac{9z}{(z+{\tilde{m}_\sigma}^2)^3}+1\right)u^2
\eeq

\beq
\beta_{u}^{q,\pi} = 2 u^2 \beta_{\kappa}^{q,\pi}
\eeq

\beq
\beta_{u}^{q,\sigma} = \frac{6 u^2}{z+{\tilde{m}_\sigma}^2} \beta_{\kappa}^{q,\sigma}
\eeq

%

%\bibliography{/Users/piotr/Dropbox/Paper_with_Piotr_2020/Tekst_21_VII/refs}{} 
%\bibliography{/Users/Pawel/Desktop/Paper_with_Piotr/Version_PJ/refs}{} 

\begin{thebibliography}{85}%
\makeatletter
\providecommand \@ifxundefined [1]{%
 \@ifx{#1\undefined}
}%
\providecommand \@ifnum [1]{%
 \ifnum #1\expandafter \@firstoftwo
 \else \expandafter \@secondoftwo
 \fi
}%
\providecommand \@ifx [1]{%
 \ifx #1\expandafter \@firstoftwo
 \else \expandafter \@secondoftwo
 \fi
}%
\providecommand \natexlab [1]{#1}%
\providecommand \enquote  [1]{``#1''}%
\providecommand \bibnamefont  [1]{#1}%
\providecommand \bibfnamefont [1]{#1}%
\providecommand \citenamefont [1]{#1}%
\providecommand \href@noop [0]{\@secondoftwo}%
\providecommand \href [0]{\begingroup \@sanitize@url \@href}%
\providecommand \@href[1]{\@@startlink{#1}\@@href}%
\providecommand \@@href[1]{\endgroup#1\@@endlink}%
\providecommand \@sanitize@url [0]{\catcode `\\12\catcode `\$12\catcode
  `\&12\catcode `\#12\catcode `\^12\catcode `\_12\catcode `\%12\relax}%
\providecommand \@@startlink[1]{}%
\providecommand \@@endlink[0]{}%
\providecommand \url  [0]{\begingroup\@sanitize@url \@url }%
\providecommand \@url [1]{\endgroup\@href {#1}{\urlprefix }}%
\providecommand \urlprefix  [0]{URL }%
\providecommand \Eprint [0]{\href }%
\providecommand \doibase [0]{http://dx.doi.org/}%
\providecommand \selectlanguage [0]{\@gobble}%
\providecommand \bibinfo  [0]{\@secondoftwo}%
\providecommand \bibfield  [0]{\@secondoftwo}%
\providecommand \translation [1]{[#1]}%
\providecommand \BibitemOpen [0]{}%
\providecommand \bibitemStop [0]{}%
\providecommand \bibitemNoStop [0]{.\EOS\space}%
\providecommand \EOS [0]{\spacefactor3000\relax}%
\providecommand \BibitemShut  [1]{\csname bibitem#1\endcsname}%
\let\auto@bib@innerbib\@empty
%</preamble>
\bibitem [{\citenamefont {Gurarie}\ and\ \citenamefont
  {Radzihovsky}(2007)}]{Radzihovsky_2007}%
  \BibitemOpen
  \bibfield  {author} {\bibinfo {author} {\bibfnamefont {V.}~\bibnamefont
  {Gurarie}}\ and\ \bibinfo {author} {\bibfnamefont {L.}~\bibnamefont
  {Radzihovsky}},\ }\href {\doibase https://doi.org/10.1016/j.aop.2006.10.009}
  {\bibfield  {journal} {\bibinfo  {journal} {Annals of Physics}\ }\textbf
  {\bibinfo {volume} {322}},\ \bibinfo {pages} {2 } (\bibinfo {year}
  {2007})}\BibitemShut {NoStop}%
\bibitem [{\citenamefont {Bloch}\ \emph {et~al.}(2008)\citenamefont {Bloch},
  \citenamefont {Dalibard},\ and\ \citenamefont {Zwerger}}]{Bloch_2008}%
  \BibitemOpen
  \bibfield  {author} {\bibinfo {author} {\bibfnamefont {I.}~\bibnamefont
  {Bloch}}, \bibinfo {author} {\bibfnamefont {J.}~\bibnamefont {Dalibard}}, \
  and\ \bibinfo {author} {\bibfnamefont {W.}~\bibnamefont {Zwerger}},\ }\href
  {\doibase 10.1103/RevModPhys.80.885} {\bibfield  {journal} {\bibinfo
  {journal} {Rev. Mod. Phys.}\ }\textbf {\bibinfo {volume} {80}},\ \bibinfo
  {pages} {885} (\bibinfo {year} {2008})}\BibitemShut {NoStop}%
\bibitem [{\citenamefont {Giorgini}\ \emph {et~al.}(2008)\citenamefont
  {Giorgini}, \citenamefont {Pitaevskii},\ and\ \citenamefont
  {Stringari}}]{Giorgini_2008}%
  \BibitemOpen
  \bibfield  {author} {\bibinfo {author} {\bibfnamefont {S.}~\bibnamefont
  {Giorgini}}, \bibinfo {author} {\bibfnamefont {L.~P.}\ \bibnamefont
  {Pitaevskii}}, \ and\ \bibinfo {author} {\bibfnamefont {S.}~\bibnamefont
  {Stringari}},\ }\href {\doibase 10.1103/RevModPhys.80.1215} {\bibfield
  {journal} {\bibinfo  {journal} {Rev. Mod. Phys.}\ }\textbf {\bibinfo {volume}
  {80}},\ \bibinfo {pages} {1215} (\bibinfo {year} {2008})}\BibitemShut
  {NoStop}%
\bibitem [{\citenamefont {Gubbels}\ and\ \citenamefont
  {Stoof}(2013)}]{Gubbels_2013}%
  \BibitemOpen
  \bibfield  {author} {\bibinfo {author} {\bibfnamefont {K.}~\bibnamefont
  {Gubbels}}\ and\ \bibinfo {author} {\bibfnamefont {H.}~\bibnamefont
  {Stoof}},\ }\href {\doibase https://doi.org/10.1016/j.physrep.2012.11.004}
  {\bibfield  {journal} {\bibinfo  {journal} {Physics Reports}\ }\textbf
  {\bibinfo {volume} {525}},\ \bibinfo {pages} {255 } (\bibinfo {year}
  {2013})}\BibitemShut {NoStop}%
\bibitem [{\citenamefont {Gross}\ and\ \citenamefont
  {Bloch}(2017)}]{Gross_2017}%
  \BibitemOpen
  \bibfield  {author} {\bibinfo {author} {\bibfnamefont {C.}~\bibnamefont
  {Gross}}\ and\ \bibinfo {author} {\bibfnamefont {I.}~\bibnamefont {Bloch}},\
  }\href {\doibase 10.1126/science.aal3837} {\bibfield  {journal} {\bibinfo
  {journal} {Science}\ }\textbf {\bibinfo {volume} {357}},\ \bibinfo {pages}
  {995} (\bibinfo {year} {2017})}\BibitemShut {NoStop}%
\bibitem [{\citenamefont {Strinati}\ \emph {et~al.}(2018)\citenamefont
  {Strinati}, \citenamefont {Pieri}, \citenamefont {Rapke}, \citenamefont
  {Schuck},\ and\ \citenamefont {Urban}}]{Strinati_2018}%
  \BibitemOpen
  \bibfield  {author} {\bibinfo {author} {\bibfnamefont {G.~C.}\ \bibnamefont
  {Strinati}}, \bibinfo {author} {\bibfnamefont {P.}~\bibnamefont {Pieri}},
  \bibinfo {author} {\bibfnamefont {G.}~\bibnamefont {Rapke}}, \bibinfo
  {author} {\bibfnamefont {P.}~\bibnamefont {Schuck}}, \ and\ \bibinfo {author}
  {\bibfnamefont {M.}~\bibnamefont {Urban}},\ }\href {\doibase
  https://doi.org/10.1016/j.physrep.2018.02.004} {\bibfield  {journal}
  {\bibinfo  {journal} {Physics Reports}\ }\textbf {\bibinfo {volume} {738}},\
  \bibinfo {pages} {1 } (\bibinfo {year} {2018})}\BibitemShut {NoStop}%
\bibitem [{\citenamefont {Cooper}\ \emph {et~al.}(2019)\citenamefont {Cooper},
  \citenamefont {Dalibard},\ and\ \citenamefont {Spielman}}]{Cooper_2019}%
  \BibitemOpen
  \bibfield  {author} {\bibinfo {author} {\bibfnamefont {N.~R.}\ \bibnamefont
  {Cooper}}, \bibinfo {author} {\bibfnamefont {J.}~\bibnamefont {Dalibard}}, \
  and\ \bibinfo {author} {\bibfnamefont {I.~B.}\ \bibnamefont {Spielman}},\
  }\href {\doibase 10.1103/RevModPhys.91.015005} {\bibfield  {journal}
  {\bibinfo  {journal} {Rev. Mod. Phys.}\ }\textbf {\bibinfo {volume} {91}},\
  \bibinfo {pages} {015005} (\bibinfo {year} {2019})}\BibitemShut {NoStop}%
\bibitem [{\citenamefont {Zwerger}(2012)}]{BCS_BEC_book}%
  \BibitemOpen
  \bibinfo {editor} {\bibfnamefont {W.}~\bibnamefont {Zwerger}},\ ed.,\
  \href@noop {} {\emph {\bibinfo {title} {The BCS-BEC Crossover and the Unitary
  Fermi Gas}}}\ (\bibinfo  {publisher} {Springer Verlag},\ \bibinfo {year}
  {2012})\BibitemShut {NoStop}%
\bibitem [{\citenamefont {Randeria}\ and\ \citenamefont
  {Taylor}(2014)}]{Randeria_2014}%
  \BibitemOpen
  \bibfield  {author} {\bibinfo {author} {\bibfnamefont {M.}~\bibnamefont
  {Randeria}}\ and\ \bibinfo {author} {\bibfnamefont {E.}~\bibnamefont
  {Taylor}},\ }\href {\doibase 10.1146/annurev-conmatphys-031113-133829}
  {\bibfield  {journal} {\bibinfo  {journal} {Annual Review of Condensed Matter
  Physics}\ }\textbf {\bibinfo {volume} {5}},\ \bibinfo {pages} {209} (\bibinfo
  {year} {2014})}\BibitemShut {NoStop}%
\bibitem [{\citenamefont {Zwierlein}\ \emph {et~al.}(2006)\citenamefont
  {Zwierlein}, \citenamefont {Schirotzek}, \citenamefont {Schunck},\ and\
  \citenamefont {Ketterle}}]{Zwierlein_2006}%
  \BibitemOpen
  \bibfield  {author} {\bibinfo {author} {\bibfnamefont {M.~W.}\ \bibnamefont
  {Zwierlein}}, \bibinfo {author} {\bibfnamefont {A.}~\bibnamefont
  {Schirotzek}}, \bibinfo {author} {\bibfnamefont {C.~H.}\ \bibnamefont
  {Schunck}}, \ and\ \bibinfo {author} {\bibfnamefont {W.}~\bibnamefont
  {Ketterle}},\ }\href {\doibase 10.1126/science.1122318} {\bibfield  {journal}
  {\bibinfo  {journal} {Science}\ }\textbf {\bibinfo {volume} {311}},\ \bibinfo
  {pages} {492} (\bibinfo {year} {2006})}\BibitemShut {NoStop}%
\bibitem [{\citenamefont {Partridge}\ \emph {et~al.}(2006)\citenamefont
  {Partridge}, \citenamefont {Li}, \citenamefont {Kamar}, \citenamefont
  {Liao},\ and\ \citenamefont {Hulet}}]{Partridge_2006}%
  \BibitemOpen
  \bibfield  {author} {\bibinfo {author} {\bibfnamefont {G.~B.}\ \bibnamefont
  {Partridge}}, \bibinfo {author} {\bibfnamefont {W.}~\bibnamefont {Li}},
  \bibinfo {author} {\bibfnamefont {R.~I.}\ \bibnamefont {Kamar}}, \bibinfo
  {author} {\bibfnamefont {Y.-a.}\ \bibnamefont {Liao}}, \ and\ \bibinfo
  {author} {\bibfnamefont {R.~G.}\ \bibnamefont {Hulet}},\ }\href {\doibase
  10.1126/science.1122876} {\bibfield  {journal} {\bibinfo  {journal}
  {Science}\ }\textbf {\bibinfo {volume} {311}},\ \bibinfo {pages} {503}
  (\bibinfo {year} {2006})}\BibitemShut {NoStop}%
\bibitem [{\citenamefont {Ketterle}\ \emph {et~al.}(2009)\citenamefont
  {Ketterle}, \citenamefont {Shin}, \citenamefont {Schirotzek},\ and\
  \citenamefont {Schunk}}]{Ketterle_2009}%
  \BibitemOpen
  \bibfield  {author} {\bibinfo {author} {\bibfnamefont {W.}~\bibnamefont
  {Ketterle}}, \bibinfo {author} {\bibfnamefont {Y.}~\bibnamefont {Shin}},
  \bibinfo {author} {\bibfnamefont {A.}~\bibnamefont {Schirotzek}}, \ and\
  \bibinfo {author} {\bibfnamefont {C.~H.}\ \bibnamefont {Schunk}},\ }\href
  {\doibase 10.1088/0953-8984/21/16/164206} {\bibfield  {journal} {\bibinfo
  {journal} {Journal of Physics: Condensed Matter}\ }\textbf {\bibinfo {volume}
  {21}},\ \bibinfo {pages} {164206} (\bibinfo {year} {2009})}\BibitemShut
  {NoStop}%
\bibitem [{\citenamefont {Wille}\ \emph {et~al.}(2008)\citenamefont {Wille},
  \citenamefont {Spiegelhalder}, \citenamefont {Kerner}, \citenamefont {Naik},
  \citenamefont {Trenkwalder}, \citenamefont {Hendl}, \citenamefont {Schreck},
  \citenamefont {Grimm}, \citenamefont {Tiecke}, \citenamefont {Walraven},
  \citenamefont {Kokkelmans}, \citenamefont {Tiesinga},\ and\ \citenamefont
  {Julienne}}]{Wille_2008}%
  \BibitemOpen
  \bibfield  {author} {\bibinfo {author} {\bibfnamefont {E.}~\bibnamefont
  {Wille}}, \bibinfo {author} {\bibfnamefont {F.~M.}\ \bibnamefont
  {Spiegelhalder}}, \bibinfo {author} {\bibfnamefont {G.}~\bibnamefont
  {Kerner}}, \bibinfo {author} {\bibfnamefont {D.}~\bibnamefont {Naik}},
  \bibinfo {author} {\bibfnamefont {A.}~\bibnamefont {Trenkwalder}}, \bibinfo
  {author} {\bibfnamefont {G.}~\bibnamefont {Hendl}}, \bibinfo {author}
  {\bibfnamefont {F.}~\bibnamefont {Schreck}}, \bibinfo {author} {\bibfnamefont
  {R.}~\bibnamefont {Grimm}}, \bibinfo {author} {\bibfnamefont {T.~G.}\
  \bibnamefont {Tiecke}}, \bibinfo {author} {\bibfnamefont {J.~T.~M.}\
  \bibnamefont {Walraven}}, \bibinfo {author} {\bibfnamefont {S.~J. J. M.~F.}\
  \bibnamefont {Kokkelmans}}, \bibinfo {author} {\bibfnamefont
  {E.}~\bibnamefont {Tiesinga}}, \ and\ \bibinfo {author} {\bibfnamefont
  {P.~S.}\ \bibnamefont {Julienne}},\ }\href {\doibase
  10.1103/PhysRevLett.100.053201} {\bibfield  {journal} {\bibinfo  {journal}
  {Phys. Rev. Lett.}\ }\textbf {\bibinfo {volume} {100}},\ \bibinfo {pages}
  {053201} (\bibinfo {year} {2008})}\BibitemShut {NoStop}%
\bibitem [{\citenamefont {Tiecke}\ \emph {et~al.}(2010)\citenamefont {Tiecke},
  \citenamefont {Goosen}, \citenamefont {Ludewig}, \citenamefont {Gensemer},
  \citenamefont {Kraft}, \citenamefont {Kokkelmans},\ and\ \citenamefont
  {Walraven}}]{Tiecke_2010}%
  \BibitemOpen
  \bibfield  {author} {\bibinfo {author} {\bibfnamefont {T.~G.}\ \bibnamefont
  {Tiecke}}, \bibinfo {author} {\bibfnamefont {M.~R.}\ \bibnamefont {Goosen}},
  \bibinfo {author} {\bibfnamefont {A.}~\bibnamefont {Ludewig}}, \bibinfo
  {author} {\bibfnamefont {S.~D.}\ \bibnamefont {Gensemer}}, \bibinfo {author}
  {\bibfnamefont {S.}~\bibnamefont {Kraft}}, \bibinfo {author} {\bibfnamefont
  {S.~J. J. M.~F.}\ \bibnamefont {Kokkelmans}}, \ and\ \bibinfo {author}
  {\bibfnamefont {J.~T.~M.}\ \bibnamefont {Walraven}},\ }\href {\doibase
  10.1103/PhysRevLett.104.053202} {\bibfield  {journal} {\bibinfo  {journal}
  {Phys. Rev. Lett.}\ }\textbf {\bibinfo {volume} {104}},\ \bibinfo {pages}
  {053202} (\bibinfo {year} {2010})}\BibitemShut {NoStop}%
\bibitem [{\citenamefont {Trenkwalder}\ \emph {et~al.}(2011)\citenamefont
  {Trenkwalder}, \citenamefont {Kohstall}, \citenamefont {Zaccanti},
  \citenamefont {Naik}, \citenamefont {Sidorov}, \citenamefont {Schreck},\ and\
  \citenamefont {Grimm}}]{Trenkwalder_2011}%
  \BibitemOpen
  \bibfield  {author} {\bibinfo {author} {\bibfnamefont {A.}~\bibnamefont
  {Trenkwalder}}, \bibinfo {author} {\bibfnamefont {C.}~\bibnamefont
  {Kohstall}}, \bibinfo {author} {\bibfnamefont {M.}~\bibnamefont {Zaccanti}},
  \bibinfo {author} {\bibfnamefont {D.}~\bibnamefont {Naik}}, \bibinfo {author}
  {\bibfnamefont {A.~I.}\ \bibnamefont {Sidorov}}, \bibinfo {author}
  {\bibfnamefont {F.}~\bibnamefont {Schreck}}, \ and\ \bibinfo {author}
  {\bibfnamefont {R.}~\bibnamefont {Grimm}},\ }\href {\doibase
  10.1103/PhysRevLett.106.115304} {\bibfield  {journal} {\bibinfo  {journal}
  {Phys. Rev. Lett.}\ }\textbf {\bibinfo {volume} {106}},\ \bibinfo {pages}
  {115304} (\bibinfo {year} {2011})}\BibitemShut {NoStop}%
\bibitem [{\citenamefont {Jag}\ \emph {et~al.}(2014)\citenamefont {Jag},
  \citenamefont {Zaccanti}, \citenamefont {Cetina}, \citenamefont {Lous},
  \citenamefont {Schreck}, \citenamefont {Grimm}, \citenamefont {Petrov},\ and\
  \citenamefont {Levinsen}}]{Jag_2014}%
  \BibitemOpen
  \bibfield  {author} {\bibinfo {author} {\bibfnamefont {M.}~\bibnamefont
  {Jag}}, \bibinfo {author} {\bibfnamefont {M.}~\bibnamefont {Zaccanti}},
  \bibinfo {author} {\bibfnamefont {M.}~\bibnamefont {Cetina}}, \bibinfo
  {author} {\bibfnamefont {R.~S.}\ \bibnamefont {Lous}}, \bibinfo {author}
  {\bibfnamefont {F.}~\bibnamefont {Schreck}}, \bibinfo {author} {\bibfnamefont
  {R.}~\bibnamefont {Grimm}}, \bibinfo {author} {\bibfnamefont {D.~S.}\
  \bibnamefont {Petrov}}, \ and\ \bibinfo {author} {\bibfnamefont
  {J.}~\bibnamefont {Levinsen}},\ }\href {\doibase
  10.1103/PhysRevLett.112.075302} {\bibfield  {journal} {\bibinfo  {journal}
  {Phys. Rev. Lett.}\ }\textbf {\bibinfo {volume} {112}},\ \bibinfo {pages}
  {075302} (\bibinfo {year} {2014})}\BibitemShut {NoStop}%
\bibitem [{\citenamefont {Ravensbergen}\ \emph {et~al.}(2018)\citenamefont
  {Ravensbergen}, \citenamefont {Corre}, \citenamefont {Soave}, \citenamefont
  {Kreyer}, \citenamefont {Kirilov},\ and\ \citenamefont
  {Grimm}}]{Ravensbergen_2018}%
  \BibitemOpen
  \bibfield  {author} {\bibinfo {author} {\bibfnamefont {C.}~\bibnamefont
  {Ravensbergen}}, \bibinfo {author} {\bibfnamefont {V.}~\bibnamefont {Corre}},
  \bibinfo {author} {\bibfnamefont {E.}~\bibnamefont {Soave}}, \bibinfo
  {author} {\bibfnamefont {M.}~\bibnamefont {Kreyer}}, \bibinfo {author}
  {\bibfnamefont {E.}~\bibnamefont {Kirilov}}, \ and\ \bibinfo {author}
  {\bibfnamefont {R.}~\bibnamefont {Grimm}},\ }\href {\doibase
  10.1103/PhysRevA.98.063624} {\bibfield  {journal} {\bibinfo  {journal} {Phys.
  Rev. A}\ }\textbf {\bibinfo {volume} {98}},\ \bibinfo {pages} {063624}
  (\bibinfo {year} {2018})}\BibitemShut {NoStop}%
\bibitem [{\citenamefont {Ravensbergen}\ \emph {et~al.}(2020)\citenamefont
  {Ravensbergen}, \citenamefont {Soave}, \citenamefont {Corre}, \citenamefont
  {Kreyer}, \citenamefont {Huang}, \citenamefont {Kirilov},\ and\ \citenamefont
  {Grimm}}]{Ravensbergen_2020}%
  \BibitemOpen
  \bibfield  {author} {\bibinfo {author} {\bibfnamefont {C.}~\bibnamefont
  {Ravensbergen}}, \bibinfo {author} {\bibfnamefont {E.}~\bibnamefont {Soave}},
  \bibinfo {author} {\bibfnamefont {V.}~\bibnamefont {Corre}}, \bibinfo
  {author} {\bibfnamefont {M.}~\bibnamefont {Kreyer}}, \bibinfo {author}
  {\bibfnamefont {B.}~\bibnamefont {Huang}}, \bibinfo {author} {\bibfnamefont
  {E.}~\bibnamefont {Kirilov}}, \ and\ \bibinfo {author} {\bibfnamefont
  {R.}~\bibnamefont {Grimm}},\ }\href {\doibase 10.1103/PhysRevLett.124.203402}
  {\bibfield  {journal} {\bibinfo  {journal} {Phys. Rev. Lett.}\ }\textbf
  {\bibinfo {volume} {124}},\ \bibinfo {pages} {203402} (\bibinfo {year}
  {2020})}\BibitemShut {NoStop}%
\bibitem [{\citenamefont {Neri}\ \emph {et~al.}(2020)\citenamefont {Neri},
  \citenamefont {Ciamei}, \citenamefont {Simonelli}, \citenamefont {Goti},
  \citenamefont {Inguscio}, \citenamefont {Trenkwalder},\ and\ \citenamefont
  {Zaccanti}}]{Neri_2020}%
  \BibitemOpen
  \bibfield  {author} {\bibinfo {author} {\bibfnamefont {E.}~\bibnamefont
  {Neri}}, \bibinfo {author} {\bibfnamefont {A.}~\bibnamefont {Ciamei}},
  \bibinfo {author} {\bibfnamefont {C.}~\bibnamefont {Simonelli}}, \bibinfo
  {author} {\bibfnamefont {I.}~\bibnamefont {Goti}}, \bibinfo {author}
  {\bibfnamefont {M.}~\bibnamefont {Inguscio}}, \bibinfo {author}
  {\bibfnamefont {A.}~\bibnamefont {Trenkwalder}}, \ and\ \bibinfo {author}
  {\bibfnamefont {M.}~\bibnamefont {Zaccanti}},\ }\href {\doibase
  10.1103/PhysRevA.101.063602} {\bibfield  {journal} {\bibinfo  {journal}
  {Phys. Rev. A}\ }\textbf {\bibinfo {volume} {101}},\ \bibinfo {pages}
  {063602} (\bibinfo {year} {2020})}\BibitemShut {NoStop}%
\bibitem [{\citenamefont {Hara}\ \emph {et~al.}(2011)\citenamefont {Hara},
  \citenamefont {Takasu}, \citenamefont {Yamaoka}, \citenamefont {Doyle},\ and\
  \citenamefont {Takahashi}}]{Hara_2011}%
  \BibitemOpen
  \bibfield  {author} {\bibinfo {author} {\bibfnamefont {H.}~\bibnamefont
  {Hara}}, \bibinfo {author} {\bibfnamefont {Y.}~\bibnamefont {Takasu}},
  \bibinfo {author} {\bibfnamefont {Y.}~\bibnamefont {Yamaoka}}, \bibinfo
  {author} {\bibfnamefont {J.~M.}\ \bibnamefont {Doyle}}, \ and\ \bibinfo
  {author} {\bibfnamefont {Y.}~\bibnamefont {Takahashi}},\ }\href {\doibase
  10.1103/PhysRevLett.106.205304} {\bibfield  {journal} {\bibinfo  {journal}
  {Phys. Rev. Lett.}\ }\textbf {\bibinfo {volume} {106}},\ \bibinfo {pages}
  {205304} (\bibinfo {year} {2011})}\BibitemShut {NoStop}%
\bibitem [{\citenamefont {Sarma}(1963)}]{Sarma_1963}%
  \BibitemOpen
  \bibfield  {author} {\bibinfo {author} {\bibfnamefont {G.}~\bibnamefont
  {Sarma}},\ }\href {\doibase https://doi.org/10.1016/0022-3697(63)90007-6}
  {\bibfield  {journal} {\bibinfo  {journal} {Journal of Physics and Chemistry
  of Solids}\ }\textbf {\bibinfo {volume} {24}},\ \bibinfo {pages} {1029 }
  (\bibinfo {year} {1963})}\BibitemShut {NoStop}%
\bibitem [{\citenamefont {Liu}\ and\ \citenamefont {Wilczek}(2003)}]{Liu_2003}%
  \BibitemOpen
  \bibfield  {author} {\bibinfo {author} {\bibfnamefont {W.~V.}\ \bibnamefont
  {Liu}}\ and\ \bibinfo {author} {\bibfnamefont {F.}~\bibnamefont {Wilczek}},\
  }\href {\doibase 10.1103/PhysRevLett.90.047002} {\bibfield  {journal}
  {\bibinfo  {journal} {Phys. Rev. Lett.}\ }\textbf {\bibinfo {volume} {90}},\
  \bibinfo {pages} {047002} (\bibinfo {year} {2003})}\BibitemShut {NoStop}%
\bibitem [{\citenamefont {Fulde}\ and\ \citenamefont
  {Ferrell}(1964)}]{fulde_superconductivity_1964}%
  \BibitemOpen
  \bibfield  {author} {\bibinfo {author} {\bibfnamefont {P.}~\bibnamefont
  {Fulde}}\ and\ \bibinfo {author} {\bibfnamefont {R.~A.}\ \bibnamefont
  {Ferrell}},\ }\href {\doibase 10.1103/PhysRev.135.A550} {\bibfield  {journal}
  {\bibinfo  {journal} {Phys. Rev.}\ }\textbf {\bibinfo {volume} {135}},\
  \bibinfo {pages} {A550} (\bibinfo {year} {1964})}\BibitemShut {NoStop}%
\bibitem [{\citenamefont {Larkin}\ and\ \citenamefont
  {Ovchinnikov}(1965)}]{larkin_nonuniform_1965}%
  \BibitemOpen
  \bibfield  {author} {\bibinfo {author} {\bibfnamefont {A.~I.}\ \bibnamefont
  {Larkin}}\ and\ \bibinfo {author} {\bibfnamefont {Y.~N.}\ \bibnamefont
  {Ovchinnikov}},\ }\href@noop {} {\bibfield  {journal} {\bibinfo  {journal}
  {Sov. Phys. JETP}\ }\textbf {\bibinfo {volume} {20}},\ \bibinfo {pages} {762}
  (\bibinfo {year} {1965})}\BibitemShut {NoStop}%
\bibitem [{\citenamefont {Iskin}\ and\ \citenamefont {S\'a~de
  Melo}(2007)}]{Iskin_2007}%
  \BibitemOpen
  \bibfield  {author} {\bibinfo {author} {\bibfnamefont {M.}~\bibnamefont
  {Iskin}}\ and\ \bibinfo {author} {\bibfnamefont {C.~A.~R.}\ \bibnamefont
  {S\'a~de Melo}},\ }\href {\doibase 10.1103/PhysRevA.76.013601} {\bibfield
  {journal} {\bibinfo  {journal} {Phys. Rev. A}\ }\textbf {\bibinfo {volume}
  {76}},\ \bibinfo {pages} {013601} (\bibinfo {year} {2007})}\BibitemShut
  {NoStop}%
\bibitem [{\citenamefont {Parish}\ \emph
  {et~al.}(2007{\natexlab{a}})\citenamefont {Parish}, \citenamefont
  {Marchetti}, \citenamefont {Lamacraft},\ and\ \citenamefont
  {Simons}}]{Parish_Nat_2007}%
  \BibitemOpen
  \bibfield  {author} {\bibinfo {author} {\bibfnamefont {M.~M.}\ \bibnamefont
  {Parish}}, \bibinfo {author} {\bibfnamefont {F.~M.}\ \bibnamefont
  {Marchetti}}, \bibinfo {author} {\bibfnamefont {A.}~\bibnamefont
  {Lamacraft}}, \ and\ \bibinfo {author} {\bibfnamefont {B.~D.}\ \bibnamefont
  {Simons}},\ }\href {\doibase 10.1038/nphys520} {\bibfield  {journal}
  {\bibinfo  {journal} {Nature Physics}\ }\textbf {\bibinfo {volume} {3}},\
  \bibinfo {pages} {124} (\bibinfo {year} {2007}{\natexlab{a}})}\BibitemShut
  {NoStop}%
\bibitem [{\citenamefont {Radzihovsky}\ and\ \citenamefont
  {Sheehy}(2010)}]{Radzihovsky_2010}%
  \BibitemOpen
  \bibfield  {author} {\bibinfo {author} {\bibfnamefont {L.}~\bibnamefont
  {Radzihovsky}}\ and\ \bibinfo {author} {\bibfnamefont {D.~E.}\ \bibnamefont
  {Sheehy}},\ }\href {\doibase 10.1088/0034-4885/73/7/076501} {\bibfield
  {journal} {\bibinfo  {journal} {Rep. Prog. Phys.}\ }\textbf {\bibinfo
  {volume} {73}},\ \bibinfo {pages} {076501} (\bibinfo {year}
  {2010})}\BibitemShut {NoStop}%
\bibitem [{\citenamefont {Klimin}\ \emph {et~al.}(2012)\citenamefont {Klimin},
  \citenamefont {Tempere},\ and\ \citenamefont {Devreese}}]{Klimin_2012}%
  \BibitemOpen
  \bibfield  {author} {\bibinfo {author} {\bibfnamefont {S.~N.}\ \bibnamefont
  {Klimin}}, \bibinfo {author} {\bibfnamefont {J.}~\bibnamefont {Tempere}}, \
  and\ \bibinfo {author} {\bibfnamefont {J.~T.}\ \bibnamefont {Devreese}},\
  }\href {\doibase 10.1088/1367-2630/14/10/103044} {\bibfield  {journal}
  {\bibinfo  {journal} {New Journal of Physics}\ }\textbf {\bibinfo {volume}
  {14}},\ \bibinfo {pages} {103044} (\bibinfo {year} {2012})}\BibitemShut
  {NoStop}%
\bibitem [{\citenamefont {Strack}\ and\ \citenamefont
  {Jakubczyk}(2014)}]{Strack_2014}%
  \BibitemOpen
  \bibfield  {author} {\bibinfo {author} {\bibfnamefont {P.}~\bibnamefont
  {Strack}}\ and\ \bibinfo {author} {\bibfnamefont {P.}~\bibnamefont
  {Jakubczyk}},\ }\href {\doibase 10.1103/PhysRevX.4.021012} {\bibfield
  {journal} {\bibinfo  {journal} {Phys. Rev. X}\ }\textbf {\bibinfo {volume}
  {4}},\ \bibinfo {pages} {021012} (\bibinfo {year} {2014})}\BibitemShut
  {NoStop}%
\bibitem [{\citenamefont {Roscher}\ \emph {et~al.}(2015)\citenamefont
  {Roscher}, \citenamefont {Braun},\ and\ \citenamefont {Drut}}]{Roscher_2015}%
  \BibitemOpen
  \bibfield  {author} {\bibinfo {author} {\bibfnamefont {D.}~\bibnamefont
  {Roscher}}, \bibinfo {author} {\bibfnamefont {J.}~\bibnamefont {Braun}}, \
  and\ \bibinfo {author} {\bibfnamefont {J.~E.}\ \bibnamefont {Drut}},\ }\href
  {\doibase 10.1103/PhysRevA.91.053611} {\bibfield  {journal} {\bibinfo
  {journal} {Phys. Rev. A}\ }\textbf {\bibinfo {volume} {91}},\ \bibinfo
  {pages} {053611} (\bibinfo {year} {2015})}\BibitemShut {NoStop}%
\bibitem [{\citenamefont {Wang}\ \emph {et~al.}(2017)\citenamefont {Wang},
  \citenamefont {Che}, \citenamefont {Zhang},\ and\ \citenamefont
  {Chen}}]{Wang_2017}%
  \BibitemOpen
  \bibfield  {author} {\bibinfo {author} {\bibfnamefont {J.}~\bibnamefont
  {Wang}}, \bibinfo {author} {\bibfnamefont {Y.}~\bibnamefont {Che}}, \bibinfo
  {author} {\bibfnamefont {L.}~\bibnamefont {Zhang}}, \ and\ \bibinfo {author}
  {\bibfnamefont {Q.}~\bibnamefont {Chen}},\ }\href {\doibase
  10.1038/srep39783} {\bibfield  {journal} {\bibinfo  {journal} {Scientific
  Reports}\ }\textbf {\bibinfo {volume} {7}},\ \bibinfo {pages} {39783}
  (\bibinfo {year} {2017})}\BibitemShut {NoStop}%
\bibitem [{\citenamefont {Parish}\ \emph
  {et~al.}(2007{\natexlab{b}})\citenamefont {Parish}, \citenamefont
  {Marchetti}, \citenamefont {Lamacraft},\ and\ \citenamefont
  {Simons}}]{Parish_2007}%
  \BibitemOpen
  \bibfield  {author} {\bibinfo {author} {\bibfnamefont {M.~M.}\ \bibnamefont
  {Parish}}, \bibinfo {author} {\bibfnamefont {F.~M.}\ \bibnamefont
  {Marchetti}}, \bibinfo {author} {\bibfnamefont {A.}~\bibnamefont
  {Lamacraft}}, \ and\ \bibinfo {author} {\bibfnamefont {B.~D.}\ \bibnamefont
  {Simons}},\ }\href {\doibase 10.1103/PhysRevLett.98.160402} {\bibfield
  {journal} {\bibinfo  {journal} {Phys. Rev. Lett.}\ }\textbf {\bibinfo
  {volume} {98}},\ \bibinfo {pages} {160402} (\bibinfo {year}
  {2007}{\natexlab{b}})}\BibitemShut {NoStop}%
\bibitem [{\citenamefont {Lamacraft}\ and\ \citenamefont
  {Marchetti}(2008)}]{Lamacraft_2008}%
  \BibitemOpen
  \bibfield  {author} {\bibinfo {author} {\bibfnamefont {A.}~\bibnamefont
  {Lamacraft}}\ and\ \bibinfo {author} {\bibfnamefont {F.~M.}\ \bibnamefont
  {Marchetti}},\ }\href {\doibase 10.1103/PhysRevB.77.014511} {\bibfield
  {journal} {\bibinfo  {journal} {Phys. Rev. B}\ }\textbf {\bibinfo {volume}
  {77}},\ \bibinfo {pages} {014511} (\bibinfo {year} {2008})}\BibitemShut
  {NoStop}%
\bibitem [{\citenamefont {Chevy}\ and\ \citenamefont
  {Mora}(2010)}]{Chevy_2010}%
  \BibitemOpen
  \bibfield  {author} {\bibinfo {author} {\bibfnamefont {F.}~\bibnamefont
  {Chevy}}\ and\ \bibinfo {author} {\bibfnamefont {C.}~\bibnamefont {Mora}},\
  }\href {\doibase 10.1088/0034-4885/73/11/112401} {\bibfield  {journal}
  {\bibinfo  {journal} {Reports on Progress in Physics}\ }\textbf {\bibinfo
  {volume} {73}},\ \bibinfo {pages} {112401} (\bibinfo {year}
  {2010})}\BibitemShut {NoStop}%
\bibitem [{\citenamefont {Zdybel}\ and\ \citenamefont
  {Jakubczyk}(2018)}]{Zdybel_2018}%
  \BibitemOpen
  \bibfield  {author} {\bibinfo {author} {\bibfnamefont {P.}~\bibnamefont
  {Zdybel}}\ and\ \bibinfo {author} {\bibfnamefont {P.}~\bibnamefont
  {Jakubczyk}},\ }\href {\doibase 10.1088/1361-648x/aacc00} {\bibfield
  {journal} {\bibinfo  {journal} {Journal of Physics: Condensed Matter}\
  }\textbf {\bibinfo {volume} {30}},\ \bibinfo {pages} {305604} (\bibinfo
  {year} {2018})}\BibitemShut {NoStop}%
\bibitem [{\citenamefont {Gubbels}\ \emph {et~al.}(2009)\citenamefont
  {Gubbels}, \citenamefont {Baarsma},\ and\ \citenamefont
  {Stoof}}]{Gubbels_2009}%
  \BibitemOpen
  \bibfield  {author} {\bibinfo {author} {\bibfnamefont {K.~B.}\ \bibnamefont
  {Gubbels}}, \bibinfo {author} {\bibfnamefont {J.~E.}\ \bibnamefont
  {Baarsma}}, \ and\ \bibinfo {author} {\bibfnamefont {H.~T.~C.}\ \bibnamefont
  {Stoof}},\ }\href {\doibase 10.1103/PhysRevLett.103.195301} {\bibfield
  {journal} {\bibinfo  {journal} {Phys. Rev. Lett.}\ }\textbf {\bibinfo
  {volume} {103}},\ \bibinfo {pages} {195301} (\bibinfo {year}
  {2009})}\BibitemShut {NoStop}%
\bibitem [{\citenamefont {Baarsma}\ \emph {et~al.}(2010)\citenamefont
  {Baarsma}, \citenamefont {Gubbels},\ and\ \citenamefont
  {Stoof}}]{baarsma_population_2010}%
  \BibitemOpen
  \bibfield  {author} {\bibinfo {author} {\bibfnamefont {J.~E.}\ \bibnamefont
  {Baarsma}}, \bibinfo {author} {\bibfnamefont {K.~B.}\ \bibnamefont
  {Gubbels}}, \ and\ \bibinfo {author} {\bibfnamefont {H.~T.~C.}\ \bibnamefont
  {Stoof}},\ }\href {\doibase 10.1103/PhysRevA.82.013624} {\bibfield  {journal}
  {\bibinfo  {journal} {Phys. Rev. A}\ }\textbf {\bibinfo {volume} {82}},\
  \bibinfo {pages} {013624} (\bibinfo {year} {2010})}\BibitemShut {NoStop}%
\bibitem [{\citenamefont {Diehl}(2002)}]{Diehl_2002}%
  \BibitemOpen
  \bibfield  {author} {\bibinfo {author} {\bibfnamefont {H.~W.}\ \bibnamefont
  {Diehl}},\ }\href@noop {} {\bibfield  {journal} {\bibinfo  {journal} {Acta
  Physica Slovaca}\ }\textbf {\bibinfo {volume} {52}},\ \bibinfo {pages} {271}
  (\bibinfo {year} {2002})}\BibitemShut {NoStop}%
\bibitem [{\citenamefont {Diehl}\ \emph {et~al.}(2003)\citenamefont {Diehl},
  \citenamefont {Shpot},\ and\ \citenamefont {Zia}}]{Diehl_2003}%
  \BibitemOpen
  \bibfield  {author} {\bibinfo {author} {\bibfnamefont {H.~W.}\ \bibnamefont
  {Diehl}}, \bibinfo {author} {\bibfnamefont {M.~A.}\ \bibnamefont {Shpot}}, \
  and\ \bibinfo {author} {\bibfnamefont {R.~K.~P.}\ \bibnamefont {Zia}},\
  }\href {\doibase 10.1103/PhysRevB.68.224415} {\bibfield  {journal} {\bibinfo
  {journal} {Phys. Rev. B}\ }\textbf {\bibinfo {volume} {68}},\ \bibinfo
  {pages} {224415} (\bibinfo {year} {2003})}\BibitemShut {NoStop}%
\bibitem [{\citenamefont {Zappal\`a}(2018)}]{Zappala_2018}%
  \BibitemOpen
  \bibfield  {author} {\bibinfo {author} {\bibfnamefont {D.}~\bibnamefont
  {Zappal\`a}},\ }\href {\doibase 10.1103/PhysRevD.98.085005} {\bibfield
  {journal} {\bibinfo  {journal} {Phys. Rev. D}\ }\textbf {\bibinfo {volume}
  {98}},\ \bibinfo {pages} {085005} (\bibinfo {year} {2018})}\BibitemShut
  {NoStop}%
\bibitem [{\citenamefont {{Essafi, K.}}\ \emph {et~al.}(2012)\citenamefont
  {{Essafi, K.}}, \citenamefont {{Kownacki, J.-P.}},\ and\ \citenamefont
  {{Mouhanna, D.}}}]{Essafi_2012}%
  \BibitemOpen
  \bibfield  {author} {\bibinfo {author} {\bibnamefont {{Essafi, K.}}},
  \bibinfo {author} {\bibnamefont {{Kownacki, J.-P.}}}, \ and\ \bibinfo
  {author} {\bibnamefont {{Mouhanna, D.}}},\ }\href {\doibase
  10.1209/0295-5075/98/51002} {\bibfield  {journal} {\bibinfo  {journal} {EPL}\
  }\textbf {\bibinfo {volume} {98}},\ \bibinfo {pages} {51002} (\bibinfo {year}
  {2012})}\BibitemShut {NoStop}%
\bibitem [{\citenamefont {Zappalà}(2017)}]{Zappala_2017}%
  \BibitemOpen
  \bibfield  {author} {\bibinfo {author} {\bibfnamefont {D.}~\bibnamefont
  {Zappalà}},\ }\href {\doibase
  https://doi.org/10.1016/j.physletb.2017.08.051} {\bibfield  {journal}
  {\bibinfo  {journal} {Physics Letters B}\ }\textbf {\bibinfo {volume}
  {773}},\ \bibinfo {pages} {213 } (\bibinfo {year} {2017})}\BibitemShut
  {NoStop}%
\bibitem [{\citenamefont {Hertz}(1976)}]{Hertz_1974}%
  \BibitemOpen
  \bibfield  {author} {\bibinfo {author} {\bibfnamefont {J.~A.}\ \bibnamefont
  {Hertz}},\ }\href {\doibase 10.1103/PhysRevB.14.1165} {\bibfield  {journal}
  {\bibinfo  {journal} {Phys. Rev. B}\ }\textbf {\bibinfo {volume} {14}},\
  \bibinfo {pages} {1165} (\bibinfo {year} {1976})}\BibitemShut {NoStop}%
\bibitem [{\citenamefont {Millis}(1993)}]{Millis_1993}%
  \BibitemOpen
  \bibfield  {author} {\bibinfo {author} {\bibfnamefont {A.~J.}\ \bibnamefont
  {Millis}},\ }\href {\doibase 10.1103/PhysRevB.48.7183} {\bibfield  {journal}
  {\bibinfo  {journal} {Phys. Rev. B}\ }\textbf {\bibinfo {volume} {48}},\
  \bibinfo {pages} {7183} (\bibinfo {year} {1993})}\BibitemShut {NoStop}%
\bibitem [{\citenamefont {Zdybel}\ and\ \citenamefont
  {Jakubczyk}(2019)}]{Zdybel_2019}%
  \BibitemOpen
  \bibfield  {author} {\bibinfo {author} {\bibfnamefont {P.}~\bibnamefont
  {Zdybel}}\ and\ \bibinfo {author} {\bibfnamefont {P.}~\bibnamefont
  {Jakubczyk}},\ }\href {\doibase 10.1103/PhysRevA.100.053622} {\bibfield
  {journal} {\bibinfo  {journal} {Phys. Rev. A}\ }\textbf {\bibinfo {volume}
  {100}},\ \bibinfo {pages} {053622} (\bibinfo {year} {2019})}\BibitemShut
  {NoStop}%
\bibitem [{\citenamefont {Belitz}\ \emph
  {et~al.}(2005{\natexlab{a}})\citenamefont {Belitz}, \citenamefont
  {Kirkpatrick},\ and\ \citenamefont {Vojta}}]{Belitz_2005}%
  \BibitemOpen
  \bibfield  {author} {\bibinfo {author} {\bibfnamefont {D.}~\bibnamefont
  {Belitz}}, \bibinfo {author} {\bibfnamefont {T.~R.}\ \bibnamefont
  {Kirkpatrick}}, \ and\ \bibinfo {author} {\bibfnamefont {T.}~\bibnamefont
  {Vojta}},\ }\href {\doibase 10.1103/RevModPhys.77.579} {\bibfield  {journal}
  {\bibinfo  {journal} {Rev. Mod. Phys.}\ }\textbf {\bibinfo {volume} {77}},\
  \bibinfo {pages} {579} (\bibinfo {year} {2005}{\natexlab{a}})}\BibitemShut
  {NoStop}%
\bibitem [{\citenamefont {Jakubczyk}\ \emph {et~al.}(2008)\citenamefont
  {Jakubczyk}, \citenamefont {Strack}, \citenamefont {Katanin},\ and\
  \citenamefont {Metzner}}]{Jakubczyk_2008}%
  \BibitemOpen
  \bibfield  {author} {\bibinfo {author} {\bibfnamefont {P.}~\bibnamefont
  {Jakubczyk}}, \bibinfo {author} {\bibfnamefont {P.}~\bibnamefont {Strack}},
  \bibinfo {author} {\bibfnamefont {A.~A.}\ \bibnamefont {Katanin}}, \ and\
  \bibinfo {author} {\bibfnamefont {W.}~\bibnamefont {Metzner}},\ }\href
  {\doibase 10.1103/PhysRevB.77.195120} {\bibfield  {journal} {\bibinfo
  {journal} {Phys. Rev. B}\ }\textbf {\bibinfo {volume} {77}},\ \bibinfo
  {pages} {195120} (\bibinfo {year} {2008})}\BibitemShut {NoStop}%
\bibitem [{\citenamefont {Altland}\ and\ \citenamefont
  {Simons}(2010)}]{Altland_book}%
  \BibitemOpen
  \bibfield  {author} {\bibinfo {author} {\bibfnamefont {A.}~\bibnamefont
  {Altland}}\ and\ \bibinfo {author} {\bibfnamefont {B.}~\bibnamefont
  {Simons}},\ }\href@noop {} {\emph {\bibinfo {title} {Condensed Matter Field
  theory}}}\ (\bibinfo  {publisher} {Cambridge University Press},\ \bibinfo
  {year} {2010})\BibitemShut {NoStop}%
\bibitem [{\citenamefont {Pieri}\ \emph {et~al.}(2004)\citenamefont {Pieri},
  \citenamefont {Pisani},\ and\ \citenamefont {Strinati}}]{Pieri_2004}%
  \BibitemOpen
  \bibfield  {author} {\bibinfo {author} {\bibfnamefont {P.}~\bibnamefont
  {Pieri}}, \bibinfo {author} {\bibfnamefont {L.}~\bibnamefont {Pisani}}, \
  and\ \bibinfo {author} {\bibfnamefont {G.~C.}\ \bibnamefont {Strinati}},\
  }\href {\doibase 10.1103/PhysRevB.70.094508} {\bibfield  {journal} {\bibinfo
  {journal} {Phys. Rev. B}\ }\textbf {\bibinfo {volume} {70}},\ \bibinfo
  {pages} {094508} (\bibinfo {year} {2004})}\BibitemShut {NoStop}%
\bibitem [{\citenamefont {Kurkjian}\ and\ \citenamefont
  {Tempere}(2017)}]{Kurkjian_2017}%
  \BibitemOpen
  \bibfield  {author} {\bibinfo {author} {\bibfnamefont {H.}~\bibnamefont
  {Kurkjian}}\ and\ \bibinfo {author} {\bibfnamefont {J.}~\bibnamefont
  {Tempere}},\ }\href {\doibase 10.1088/1367-2630/aa969b} {\bibfield  {journal}
  {\bibinfo  {journal} {New J. Phys.}\ }\textbf {\bibinfo {volume} {19}},\
  \bibinfo {pages} {113045} (\bibinfo {year} {2017})}\BibitemShut {NoStop}%
\bibitem [{\citenamefont {Klimin}\ \emph {et~al.}(2019)\citenamefont {Klimin},
  \citenamefont {Tempere},\ and\ \citenamefont {Kurkjian}}]{Klimin_2019}%
  \BibitemOpen
  \bibfield  {author} {\bibinfo {author} {\bibfnamefont {S.~N.}\ \bibnamefont
  {Klimin}}, \bibinfo {author} {\bibfnamefont {J.}~\bibnamefont {Tempere}}, \
  and\ \bibinfo {author} {\bibfnamefont {H.}~\bibnamefont {Kurkjian}},\ }\href
  {\doibase 10.1103/PhysRevA.100.063634} {\bibfield  {journal} {\bibinfo
  {journal} {Phys. Rev. A}\ }\textbf {\bibinfo {volume} {100}},\ \bibinfo
  {pages} {063634} (\bibinfo {year} {2019})}\BibitemShut {NoStop}%
\bibitem [{\citenamefont {Nagaosa}(1999)}]{Nagaosa_book}%
  \BibitemOpen
  \bibfield  {author} {\bibinfo {author} {\bibfnamefont {N.}~\bibnamefont
  {Nagaosa}},\ }\href@noop {} {\emph {\bibinfo {title} {Quantum Field Theory in
  Strongly Correlated Electronic Systems}}}\ (\bibinfo  {publisher} {Springer
  Verlag},\ \bibinfo {year} {1999})\BibitemShut {NoStop}%
\bibitem [{\citenamefont {Continentino}(2017)}]{Continentino_book}%
  \BibitemOpen
  \bibfield  {author} {\bibinfo {author} {\bibfnamefont {M.}~\bibnamefont
  {Continentino}},\ }\href@noop {} {\emph {\bibinfo {title} {Quantum Scaling in
  Many-Body Systems}}},\ \bibinfo {edition} {2nd}\ ed.\ (\bibinfo  {publisher}
  {Cambridge University Press},\ \bibinfo {year} {2017})\BibitemShut {NoStop}%
\bibitem [{\citenamefont {L\"ohneysen}\ \emph {et~al.}(2007)\citenamefont
  {L\"ohneysen}, \citenamefont {Rosch}, \citenamefont {Vojta},\ and\
  \citenamefont {W\"olfle}}]{Lohneysen_2007}%
  \BibitemOpen
  \bibfield  {author} {\bibinfo {author} {\bibfnamefont {H.~v.}\ \bibnamefont
  {L\"ohneysen}}, \bibinfo {author} {\bibfnamefont {A.}~\bibnamefont {Rosch}},
  \bibinfo {author} {\bibfnamefont {M.}~\bibnamefont {Vojta}}, \ and\ \bibinfo
  {author} {\bibfnamefont {P.}~\bibnamefont {W\"olfle}},\ }\href {\doibase
  10.1103/RevModPhys.79.1015} {\bibfield  {journal} {\bibinfo  {journal} {Rev.
  Mod. Phys.}\ }\textbf {\bibinfo {volume} {79}},\ \bibinfo {pages} {1015}
  (\bibinfo {year} {2007})}\BibitemShut {NoStop}%
\bibitem [{\citenamefont {Pistolesi}\ \emph {et~al.}(2004)\citenamefont
  {Pistolesi}, \citenamefont {Castellani}, \citenamefont {Di~Castro},\ and\
  \citenamefont {Strinati}}]{Pistolesi_2004}%
  \BibitemOpen
  \bibfield  {author} {\bibinfo {author} {\bibfnamefont {F.}~\bibnamefont
  {Pistolesi}}, \bibinfo {author} {\bibfnamefont {C.}~\bibnamefont
  {Castellani}}, \bibinfo {author} {\bibfnamefont {C.}~\bibnamefont
  {Di~Castro}}, \ and\ \bibinfo {author} {\bibfnamefont {G.~C.}\ \bibnamefont
  {Strinati}},\ }\href {\doibase 10.1103/PhysRevB.69.024513} {\bibfield
  {journal} {\bibinfo  {journal} {Phys. Rev. B}\ }\textbf {\bibinfo {volume}
  {69}},\ \bibinfo {pages} {024513} (\bibinfo {year} {2004})}\BibitemShut
  {NoStop}%
\bibitem [{\citenamefont {Obert}\ \emph {et~al.}(2013)\citenamefont {Obert},
  \citenamefont {Husemann},\ and\ \citenamefont {Metzner}}]{Obert_2013}%
  \BibitemOpen
  \bibfield  {author} {\bibinfo {author} {\bibfnamefont {B.}~\bibnamefont
  {Obert}}, \bibinfo {author} {\bibfnamefont {C.}~\bibnamefont {Husemann}}, \
  and\ \bibinfo {author} {\bibfnamefont {W.}~\bibnamefont {Metzner}},\ }\href
  {\doibase 10.1103/PhysRevB.88.144508} {\bibfield  {journal} {\bibinfo
  {journal} {Phys. Rev. B}\ }\textbf {\bibinfo {volume} {88}},\ \bibinfo
  {pages} {144508} (\bibinfo {year} {2013})}\BibitemShut {NoStop}%
\bibitem [{\citenamefont {Boettcher}\ \emph {et~al.}(2015)\citenamefont
  {Boettcher}, \citenamefont {Braun}, \citenamefont {Herbst}, \citenamefont
  {Pawlowski}, \citenamefont {Roscher},\ and\ \citenamefont
  {Wetterich}}]{Boettcher_2015}%
  \BibitemOpen
  \bibfield  {author} {\bibinfo {author} {\bibfnamefont {I.}~\bibnamefont
  {Boettcher}}, \bibinfo {author} {\bibfnamefont {J.}~\bibnamefont {Braun}},
  \bibinfo {author} {\bibfnamefont {T.~K.}\ \bibnamefont {Herbst}}, \bibinfo
  {author} {\bibfnamefont {J.~M.}\ \bibnamefont {Pawlowski}}, \bibinfo {author}
  {\bibfnamefont {D.}~\bibnamefont {Roscher}}, \ and\ \bibinfo {author}
  {\bibfnamefont {C.}~\bibnamefont {Wetterich}},\ }\href {\doibase
  10.1103/PhysRevA.91.013610} {\bibfield  {journal} {\bibinfo  {journal} {Phys.
  Rev. A}\ }\textbf {\bibinfo {volume} {91}},\ \bibinfo {pages} {013610}
  (\bibinfo {year} {2015})}\BibitemShut {NoStop}%
\bibitem [{\citenamefont {Nozi{\`e}res}\ and\ \citenamefont
  {Schmitt-Rink}(1985)}]{Nozires_1985}%
  \BibitemOpen
  \bibfield  {author} {\bibinfo {author} {\bibfnamefont {P.}~\bibnamefont
  {Nozi{\`e}res}}\ and\ \bibinfo {author} {\bibfnamefont {S.}~\bibnamefont
  {Schmitt-Rink}},\ }\href {\doibase 10.1007/BF00683774} {\bibfield  {journal}
  {\bibinfo  {journal} {J. Low Temp. Phys.}\ }\textbf {\bibinfo {volume}
  {59}},\ \bibinfo {pages} {195} (\bibinfo {year} {1985})}\BibitemShut
  {NoStop}%
\bibitem [{\citenamefont {Liu}\ and\ \citenamefont {Hu}(2006)}]{Liu_2006}%
  \BibitemOpen
  \bibfield  {author} {\bibinfo {author} {\bibfnamefont {X.-J.}\ \bibnamefont
  {Liu}}\ and\ \bibinfo {author} {\bibfnamefont {H.}~\bibnamefont {Hu}},\
  }\href {\doibase 10.1209/epl/i2006-10136-3} {\bibfield  {journal} {\bibinfo
  {journal} {Europhysics Letters ({EPL})}\ }\textbf {\bibinfo {volume} {75}},\
  \bibinfo {pages} {364} (\bibinfo {year} {2006})}\BibitemShut {NoStop}%
\bibitem [{\citenamefont {Stoof}\ \emph {et~al.}(2009)\citenamefont {Stoof},
  \citenamefont {Dickerscheid},\ and\ \citenamefont {Gubbels}}]{Stoof_book}%
  \BibitemOpen
  \bibfield  {author} {\bibinfo {author} {\bibfnamefont {H.~T.~C.}\
  \bibnamefont {Stoof}}, \bibinfo {author} {\bibfnamefont {D.~B.~M.}\
  \bibnamefont {Dickerscheid}}, \ and\ \bibinfo {author} {\bibfnamefont
  {K.}~\bibnamefont {Gubbels}},\ }\href {\doibase 10.1017/CBO9780511789984}
  {\emph {\bibinfo {title} {Ultracold Quantum Fields}}}\ (\bibinfo  {publisher}
  {Springer Verlag},\ \bibinfo {year} {2009})\BibitemShut {NoStop}%
\bibitem [{\citenamefont {Jishi}(2013)}]{Jishi_book}%
  \BibitemOpen
  \bibfield  {author} {\bibinfo {author} {\bibfnamefont {R.~A.}\ \bibnamefont
  {Jishi}},\ }\href {\doibase 10.1017/CBO9781139177771} {\emph {\bibinfo
  {title} {Feynman Diagram Techniques in Condensed Matter Physics}}}\ (\bibinfo
   {publisher} {Cambridge University Press},\ \bibinfo {year}
  {2013})\BibitemShut {NoStop}%
\bibitem [{\citenamefont {Ramazashvili}(1999)}]{Ramazashvili_1999}%
  \BibitemOpen
  \bibfield  {author} {\bibinfo {author} {\bibfnamefont {R.}~\bibnamefont
  {Ramazashvili}},\ }\href {\doibase 10.1103/PhysRevB.60.7314} {\bibfield
  {journal} {\bibinfo  {journal} {Phys. Rev. B}\ }\textbf {\bibinfo {volume}
  {60}},\ \bibinfo {pages} {7314} (\bibinfo {year} {1999})}\BibitemShut
  {NoStop}%
\bibitem [{\citenamefont {Continentino}(2004)}]{Continentino_2004}%
  \BibitemOpen
  \bibfield  {author} {\bibinfo {author} {\bibfnamefont {M.~A.}\ \bibnamefont
  {Continentino}},\ }\href {\doibase
  https://doi.org/10.1016/j.jmmm.2003.11.097} {\bibfield  {journal} {\bibinfo
  {journal} {Journal of Magnetism and Magnetic Materials}\ }\textbf {\bibinfo
  {volume} {272-276}},\ \bibinfo {pages} {231 } (\bibinfo {year} {2004})},\
  \bibinfo {note} {proceedings of the International Conference on Magnetism
  (ICM 2003)}\BibitemShut {NoStop}%
\bibitem [{\citenamefont {Radzihovsky}(2011)}]{Radzihovsky_2011}%
  \BibitemOpen
  \bibfield  {author} {\bibinfo {author} {\bibfnamefont {L.}~\bibnamefont
  {Radzihovsky}},\ }\href {\doibase 10.1103/PhysRevA.84.023611} {\bibfield
  {journal} {\bibinfo  {journal} {Phys. Rev. A}\ }\textbf {\bibinfo {volume}
  {84}},\ \bibinfo {pages} {023611} (\bibinfo {year} {2011})}\BibitemShut
  {NoStop}%
\bibitem [{\citenamefont {Jakubczyk}(2017)}]{Jakubczyk_2017}%
  \BibitemOpen
  \bibfield  {author} {\bibinfo {author} {\bibfnamefont {P.}~\bibnamefont
  {Jakubczyk}},\ }\href {\doibase 10.1103/PhysRevA.95.063626} {\bibfield
  {journal} {\bibinfo  {journal} {Phys. Rev. A}\ }\textbf {\bibinfo {volume}
  {95}},\ \bibinfo {pages} {063626} (\bibinfo {year} {2017})}\BibitemShut
  {NoStop}%
\bibitem [{\citenamefont {Wang}\ \emph {et~al.}(2018)\citenamefont {Wang},
  \citenamefont {Che}, \citenamefont {Zhang},\ and\ \citenamefont
  {Chen}}]{Wang_2018}%
  \BibitemOpen
  \bibfield  {author} {\bibinfo {author} {\bibfnamefont {J.}~\bibnamefont
  {Wang}}, \bibinfo {author} {\bibfnamefont {Y.}~\bibnamefont {Che}}, \bibinfo
  {author} {\bibfnamefont {L.}~\bibnamefont {Zhang}}, \ and\ \bibinfo {author}
  {\bibfnamefont {Q.}~\bibnamefont {Chen}},\ }\href {\doibase
  10.1103/PhysRevB.97.134513} {\bibfield  {journal} {\bibinfo  {journal} {Phys.
  Rev. B}\ }\textbf {\bibinfo {volume} {97}},\ \bibinfo {pages} {134513}
  (\bibinfo {year} {2018})}\BibitemShut {NoStop}%
\bibitem [{\citenamefont {Belitz}\ \emph
  {et~al.}(2005{\natexlab{b}})\citenamefont {Belitz}, \citenamefont
  {Kirkpatrick},\ and\ \citenamefont {Rollb\"uhler}}]{Belitz_2005_2}%
  \BibitemOpen
  \bibfield  {author} {\bibinfo {author} {\bibfnamefont {D.}~\bibnamefont
  {Belitz}}, \bibinfo {author} {\bibfnamefont {T.~R.}\ \bibnamefont
  {Kirkpatrick}}, \ and\ \bibinfo {author} {\bibfnamefont {J.}~\bibnamefont
  {Rollb\"uhler}},\ }\href {\doibase 10.1103/PhysRevLett.94.247205} {\bibfield
  {journal} {\bibinfo  {journal} {Phys. Rev. Lett.}\ }\textbf {\bibinfo
  {volume} {94}},\ \bibinfo {pages} {247205} (\bibinfo {year}
  {2005}{\natexlab{b}})}\BibitemShut {NoStop}%
\bibitem [{\citenamefont {Halperin}\ \emph {et~al.}(1974)\citenamefont
  {Halperin}, \citenamefont {Lubensky},\ and\ \citenamefont
  {Ma}}]{Halperin_1974}%
  \BibitemOpen
  \bibfield  {author} {\bibinfo {author} {\bibfnamefont {B.~I.}\ \bibnamefont
  {Halperin}}, \bibinfo {author} {\bibfnamefont {T.~C.}\ \bibnamefont
  {Lubensky}}, \ and\ \bibinfo {author} {\bibfnamefont {S.-k.}\ \bibnamefont
  {Ma}},\ }\href {\doibase 10.1103/PhysRevLett.32.292} {\bibfield  {journal}
  {\bibinfo  {journal} {Phys. Rev. Lett.}\ }\textbf {\bibinfo {volume} {32}},\
  \bibinfo {pages} {292} (\bibinfo {year} {1974})}\BibitemShut {NoStop}%
\bibitem [{\citenamefont {Li}\ \emph {et~al.}(2009)\citenamefont {Li},
  \citenamefont {Belitz},\ and\ \citenamefont {Toner}}]{Li_2009}%
  \BibitemOpen
  \bibfield  {author} {\bibinfo {author} {\bibfnamefont {Q.}~\bibnamefont
  {Li}}, \bibinfo {author} {\bibfnamefont {D.}~\bibnamefont {Belitz}}, \ and\
  \bibinfo {author} {\bibfnamefont {J.}~\bibnamefont {Toner}},\ }\href
  {\doibase 10.1103/PhysRevB.79.054514} {\bibfield  {journal} {\bibinfo
  {journal} {Phys. Rev. B}\ }\textbf {\bibinfo {volume} {79}},\ \bibinfo
  {pages} {054514} (\bibinfo {year} {2009})}\BibitemShut {NoStop}%
\bibitem [{\citenamefont {Wetterich}(1993)}]{Wetterich_1993}%
  \BibitemOpen
  \bibfield  {author} {\bibinfo {author} {\bibfnamefont {C.}~\bibnamefont
  {Wetterich}},\ }\href {\doibase https://doi.org/10.1016/0370-2693(93)90726-X}
  {\bibfield  {journal} {\bibinfo  {journal} {Physics Letters B}\ }\textbf
  {\bibinfo {volume} {301}},\ \bibinfo {pages} {90 } (\bibinfo {year}
  {1993})}\BibitemShut {NoStop}%
\bibitem [{\citenamefont {Strack}\ and\ \citenamefont
  {Jakubczyk}(2009)}]{Strack_2009}%
  \BibitemOpen
  \bibfield  {author} {\bibinfo {author} {\bibfnamefont {P.}~\bibnamefont
  {Strack}}\ and\ \bibinfo {author} {\bibfnamefont {P.}~\bibnamefont
  {Jakubczyk}},\ }\href {\doibase 10.1103/PhysRevB.80.085108} {\bibfield
  {journal} {\bibinfo  {journal} {Phys. Rev. B}\ }\textbf {\bibinfo {volume}
  {80}},\ \bibinfo {pages} {085108} (\bibinfo {year} {2009})}\BibitemShut
  {NoStop}%
\bibitem [{\citenamefont {L\'eonard}\ and\ \citenamefont
  {Delamotte}(2015)}]{Leonard_2015}%
  \BibitemOpen
  \bibfield  {author} {\bibinfo {author} {\bibfnamefont {F.}~\bibnamefont
  {L\'eonard}}\ and\ \bibinfo {author} {\bibfnamefont {B.}~\bibnamefont
  {Delamotte}},\ }\href {\doibase 10.1103/PhysRevLett.115.200601} {\bibfield
  {journal} {\bibinfo  {journal} {Phys. Rev. Lett.}\ }\textbf {\bibinfo
  {volume} {115}},\ \bibinfo {pages} {200601} (\bibinfo {year}
  {2015})}\BibitemShut {NoStop}%
\bibitem [{\citenamefont {Lammers}\ \emph {et~al.}(2016)\citenamefont
  {Lammers}, \citenamefont {Boettcher},\ and\ \citenamefont
  {Wetterich}}]{Lammers_2016}%
  \BibitemOpen
  \bibfield  {author} {\bibinfo {author} {\bibfnamefont {S.}~\bibnamefont
  {Lammers}}, \bibinfo {author} {\bibfnamefont {I.}~\bibnamefont {Boettcher}},
  \ and\ \bibinfo {author} {\bibfnamefont {C.}~\bibnamefont {Wetterich}},\
  }\href {\doibase 10.1103/PhysRevA.93.063631} {\bibfield  {journal} {\bibinfo
  {journal} {Phys. Rev. A}\ }\textbf {\bibinfo {volume} {93}},\ \bibinfo
  {pages} {063631} (\bibinfo {year} {2016})}\BibitemShut {NoStop}%
\bibitem [{\citenamefont {Debelhoir}\ and\ \citenamefont
  {Dupuis}(2016)}]{Debelhoir_2016}%
  \BibitemOpen
  \bibfield  {author} {\bibinfo {author} {\bibfnamefont {T.}~\bibnamefont
  {Debelhoir}}\ and\ \bibinfo {author} {\bibfnamefont {N.}~\bibnamefont
  {Dupuis}},\ }\href {\doibase 10.1103/PhysRevA.93.023642} {\bibfield
  {journal} {\bibinfo  {journal} {Phys. Rev. A}\ }\textbf {\bibinfo {volume}
  {93}},\ \bibinfo {pages} {023642} (\bibinfo {year} {2016})}\BibitemShut
  {NoStop}%
\bibitem [{\citenamefont {Ran\ifmmode~\mbox{\c{c}}\else \c{c}\fi{}on}\ and\
  \citenamefont {Dupuis}(2017)}]{Rancon_2017}%
  \BibitemOpen
  \bibfield  {author} {\bibinfo {author} {\bibfnamefont {A.}~\bibnamefont
  {Ran\ifmmode~\mbox{\c{c}}\else \c{c}\fi{}on}}\ and\ \bibinfo {author}
  {\bibfnamefont {N.}~\bibnamefont {Dupuis}},\ }\href {\doibase
  10.1103/PhysRevB.96.214512} {\bibfield  {journal} {\bibinfo  {journal} {Phys.
  Rev. B}\ }\textbf {\bibinfo {volume} {96}},\ \bibinfo {pages} {214512}
  (\bibinfo {year} {2017})}\BibitemShut {NoStop}%
\bibitem [{\citenamefont {Chlebicki}\ and\ \citenamefont
  {Jakubczyk}(2019)}]{Chlebicki_2019}%
  \BibitemOpen
  \bibfield  {author} {\bibinfo {author} {\bibfnamefont {A.}~\bibnamefont
  {Chlebicki}}\ and\ \bibinfo {author} {\bibfnamefont {P.}~\bibnamefont
  {Jakubczyk}},\ }\href {\doibase 10.1103/PhysRevE.100.052106} {\bibfield
  {journal} {\bibinfo  {journal} {Phys. Rev. E}\ }\textbf {\bibinfo {volume}
  {100}},\ \bibinfo {pages} {052106} (\bibinfo {year} {2019})}\BibitemShut
  {NoStop}%
\bibitem [{\citenamefont {Berges}\ \emph {et~al.}(2002)\citenamefont {Berges},
  \citenamefont {Tetradis},\ and\ \citenamefont {Wetterich}}]{Berges_2002}%
  \BibitemOpen
  \bibfield  {author} {\bibinfo {author} {\bibfnamefont {J.}~\bibnamefont
  {Berges}}, \bibinfo {author} {\bibfnamefont {N.}~\bibnamefont {Tetradis}}, \
  and\ \bibinfo {author} {\bibfnamefont {C.}~\bibnamefont {Wetterich}},\ }\href
  {\doibase https://doi.org/10.1016/S0370-1573(01)00098-9} {\bibfield
  {journal} {\bibinfo  {journal} {Physics Reports}\ }\textbf {\bibinfo {volume}
  {363}},\ \bibinfo {pages} {223 } (\bibinfo {year} {2002})}\BibitemShut
  {NoStop}%
\bibitem [{\citenamefont {Kopietz}\ \emph {et~al.}(2010)\citenamefont
  {Kopietz}, \citenamefont {Bartosch},\ and\ \citenamefont
  {Sch\"utz}}]{Kopietz_book}%
  \BibitemOpen
  \bibfield  {author} {\bibinfo {author} {\bibfnamefont {P.}~\bibnamefont
  {Kopietz}}, \bibinfo {author} {\bibfnamefont {L.}~\bibnamefont {Bartosch}}, \
  and\ \bibinfo {author} {\bibfnamefont {F.}~\bibnamefont {Sch\"utz}},\
  }\href@noop {} {\emph {\bibinfo {title} {Introduction to the Functional
  Renormalization Group}}}\ (\bibinfo  {publisher} {Springer Verlag},\ \bibinfo
  {year} {2010})\BibitemShut {NoStop}%
\bibitem [{\citenamefont {Polonyi}\ and\ \citenamefont
  {Schwenk}(2012)}]{RG_book}%
  \BibitemOpen
  \bibinfo {editor} {\bibfnamefont {J.}~\bibnamefont {Polonyi}}\ and\ \bibinfo
  {editor} {\bibfnamefont {A.}~\bibnamefont {Schwenk}},\ eds.,\ \href@noop {}
  {\emph {\bibinfo {title} {Renormalization Group and Effective Field Theory
  Approaches to Many-Body Systems}}}\ (\bibinfo  {publisher} {Springer
  Verlag},\ \bibinfo {year} {2012})\BibitemShut {NoStop}%
\bibitem [{\citenamefont {Dupuis}\ \emph {et~al.}(2020)\citenamefont {Dupuis},
  \citenamefont {Canet}, \citenamefont {Eichhorn}, \citenamefont {Metzner},
  \citenamefont {Pawlowski}, \citenamefont {Tissier},\ and\ \citenamefont
  {Wschebor}}]{Dupuis_2020}%
  \BibitemOpen
  \bibfield  {author} {\bibinfo {author} {\bibfnamefont {N.}~\bibnamefont
  {Dupuis}}, \bibinfo {author} {\bibfnamefont {L.}~\bibnamefont {Canet}},
  \bibinfo {author} {\bibfnamefont {A.}~\bibnamefont {Eichhorn}}, \bibinfo
  {author} {\bibfnamefont {W.}~\bibnamefont {Metzner}}, \bibinfo {author}
  {\bibfnamefont {J.~M.}\ \bibnamefont {Pawlowski}}, \bibinfo {author}
  {\bibfnamefont {M.}~\bibnamefont {Tissier}}, \ and\ \bibinfo {author}
  {\bibfnamefont {N.}~\bibnamefont {Wschebor}},\ }\href@noop {} {\  (\bibinfo
  {year} {2020})},\ \Eprint {http://arxiv.org/abs/2006.04853} {arXiv:2006.04853
  [cond-mat.stat-mech]} \BibitemShut {NoStop}%
\bibitem [{\citenamefont {Jakubczyk}(2009)}]{Jakubczyk_2009_phi6}%
  \BibitemOpen
  \bibfield  {author} {\bibinfo {author} {\bibfnamefont {P.}~\bibnamefont
  {Jakubczyk}},\ }\href {\doibase 10.1103/PhysRevB.79.125115} {\bibfield
  {journal} {\bibinfo  {journal} {Phys. Rev. B}\ }\textbf {\bibinfo {volume}
  {79}},\ \bibinfo {pages} {125115} (\bibinfo {year} {2009})}\BibitemShut
  {NoStop}%
\bibitem [{\citenamefont {Jakubczyk}\ \emph {et~al.}(2010)\citenamefont
  {Jakubczyk}, \citenamefont {Bauer},\ and\ \citenamefont
  {Metzner}}]{Jakubczyk_2010}%
  \BibitemOpen
  \bibfield  {author} {\bibinfo {author} {\bibfnamefont {P.}~\bibnamefont
  {Jakubczyk}}, \bibinfo {author} {\bibfnamefont {J.}~\bibnamefont {Bauer}}, \
  and\ \bibinfo {author} {\bibfnamefont {W.}~\bibnamefont {Metzner}},\ }\href
  {\doibase 10.1103/PhysRevB.82.045103} {\bibfield  {journal} {\bibinfo
  {journal} {Phys. Rev. B}\ }\textbf {\bibinfo {volume} {82}},\ \bibinfo
  {pages} {045103} (\bibinfo {year} {2010})}\BibitemShut {NoStop}%
\bibitem [{\citenamefont {Litim}(2001)}]{Litim_2001}%
  \BibitemOpen
  \bibfield  {author} {\bibinfo {author} {\bibfnamefont {D.~F.}\ \bibnamefont
  {Litim}},\ }\href {\doibase 10.1103/PhysRevD.64.105007} {\bibfield  {journal}
  {\bibinfo  {journal} {Phys. Rev. D}\ }\textbf {\bibinfo {volume} {64}},\
  \bibinfo {pages} {105007} (\bibinfo {year} {2001})}\BibitemShut {NoStop}%
\bibitem [{\citenamefont {Sachdev}(2011)}]{Sachdev_book}%
  \BibitemOpen
  \bibfield  {author} {\bibinfo {author} {\bibfnamefont {S.}~\bibnamefont
  {Sachdev}},\ }\href {\doibase 10.1017/CBO9780511973765} {\emph {\bibinfo
  {title} {Quantum Phase Transitions}}},\ \bibinfo {edition} {2nd}\ ed.\
  (\bibinfo  {publisher} {Cambridge University Press},\ \bibinfo {year}
  {2011})\BibitemShut {NoStop}%
\bibitem [{\citenamefont {Chin}\ \emph {et~al.}(2010)\citenamefont {Chin},
  \citenamefont {Grimm}, \citenamefont {Julienne},\ and\ \citenamefont
  {Tiesinga}}]{Chin_2010}%
  \BibitemOpen
  \bibfield  {author} {\bibinfo {author} {\bibfnamefont {C.}~\bibnamefont
  {Chin}}, \bibinfo {author} {\bibfnamefont {R.}~\bibnamefont {Grimm}},
  \bibinfo {author} {\bibfnamefont {P.}~\bibnamefont {Julienne}}, \ and\
  \bibinfo {author} {\bibfnamefont {E.}~\bibnamefont {Tiesinga}},\ }\href
  {\doibase 10.1103/RevModPhys.82.1225} {\bibfield  {journal} {\bibinfo
  {journal} {Rev. Mod. Phys.}\ }\textbf {\bibinfo {volume} {82}},\ \bibinfo
  {pages} {1225} (\bibinfo {year} {2010})}\BibitemShut {NoStop}%
\end{thebibliography}
%\bibliographystyle{apsrev4-1}

\end{document}